\documentclass[onecolumn,preprintnumbers,amsmath,amssymb]{revtex4}

\usepackage{graphicx}% Include figure files
\usepackage{color}
\usepackage{bm}% bold math

\usepackage{hyperref}

\newcommand{\ve}[1][K]{\mathbf{#1}}

\begin{document}

\title{Kinetics of rare events for  non-Markovian  stationary processes and application to polymer dynamics}

\author{N. Levernier$^1$, O. B\'enichou$^2$, R. Voituriez$^{2,3}$, T. Gu\'erin$^{4}$}

\affiliation{$^{1}$ NCCR Chemical Biology, Departments of Biochemistry and Theoretical Physics, University of Geneva, Geneva, Switzerland}
\affiliation{$^{2}$ Laboratoire de Physique Th\'eorique de la Mati\`ere Condens\'ee, CNRS/Sorbonne University, 
 4 Place Jussieu, 75005 Paris, France}
 \affiliation{$^{3}$ Laboratoire Jean Perrin, CNRS/Sorbonne University, 
 4 Place Jussieu, 75005 Paris, France }
\affiliation{$^{4}$Laboratoire Ondes et Mati\`ere d'Aquitaine, CNRS/University of Bordeaux, F-33400 Talence, France}

\begin{abstract}
How much time does it take for a fluctuating system, such as a polymer chain,  to reach a  target configuration that is rarely visited -- typically because of a high energy cost ? This question generally amounts to the determination of the  first-passage time statistics to a  target zone in phase space with lower occupation probability. Here, we present an analytical  method to determine the mean     first-passage time of a generic non-Markovian random walker  to a rarely visited threshold, which  goes beyond existing weak-noise theories. We apply our method to polymer systems, to determine (i) the first time for a flexible polymer to reach a large extension, and (ii) the first closure time of a stiff inextensible wormlike  chain. Our results are in excellent agreement with numerical simulations and provide explicit asymptotic laws for the mean first-passage times to rarely visited configurations.
 \end{abstract}

\maketitle

The first-passage time (FPT) quantifies the time required  for a random walker to reach a  ``target'' point~\cite{Redner:2001a,metzler2014first,pal2017first,benichou2010optimal,vaccario2015first,ReviewBray,godec2016first,godec2016universal,Condamin2007,grebenkov2016universal}, with applications in contexts as varied as finance, biophysics, search processes or reactions kinetics~\cite{metzler2014first}. 
In the case of systems with many internal degrees of freedom, such as polymers or membranes, the dynamics of a single degree of freedom, e.g. a reaction coordinate, is typically non-Markovian (i.e. displays memory effects), which significantly complexifies the theoretical description of their first-passage properties~\cite{ReviewBray,VanKampen1992}. 

Generally speaking, one can distinguish between two classes of first-passage problems. First, the search for the target by the random walker can be limited by an ``entropic'' cost, such as in the case of a target located  in a large confined domain, which has been the subject of many recent studies, both for Markovian~\cite{Benichou2008,grebenkov2016universal,Schuss2007,Condamin2007,godec2016first,godec2016universal}, and non-Markovian~\cite{ReviewBray,Guerin2012a,guerin2016mean} random walks.  Second, a very different class of problems is the search of rarely visited configurations, i.e. limited by a high energy cost (or  quasipotential cost in non-equilibrium systems~\cite{Maier1992,Freidlin1984,delacruz2018minimum}). Such problem is 
 the cornerstone of reaction rate theory~\cite{hanggi1990reaction,pollak2005reaction}, but is also crucial in situations as varied  as
  population~\cite{kamenev2008colored} or disease~\cite{dykman2008disease} extinction, bond rupture~\cite{merkel1999energy,hummer2003kinetics,bullerjahn2014theory}, adhesion~\cite{jeppesen2001impact}, stock market crashes~\cite{Bouchaud1998}, or extreme heat waves in climate models~\cite{ragone2018computation}.  
The kinetics of rare events have been intensively  investigated, and explicit expressions have been proposed 
for the noise induced escape time from attraction domains in the weak noise limit~\cite{Kramers1940,schuss2009theory,bouchet2016generalisation,grote1980stable,hanggi1982thermally,hanggi1990reaction,meerson2016macroscopic}. However, for non-Markovian processes, existing approaches fail to predict quantitatively the first-passage time to a  generic rarely visited target, such as a threshold for a reaction coordinate. For example, in the context of large deviation kinetics of flexible polymers, it has recently~\cite{cao2015large} been noted that standard weak noise theories (to be defined below) lead to erroneous scalings for the mean FPT.   
  
In this Letter, we investigate the impact of memory effects on the mean time a continuous non-Markovian (possibly non Gaussian) variable  $r(t)$ takes to reach a given rarely visited threshold. We show that memory effects can be accounted for by characterizing the trajectory followed by $r(t)$ in the future of the FPT, which generalizes  a recent theoretical approach restricted to unbiased Gaussian processes~\cite{guerin2016mean}.  %The main effect is that, after the FPT, the random walker returns to its stable position \textit{infinitely faster} (i.e. with different scalings) than following an equilibrium distribution. 
%This effect can modify the kinetics by more than one order of magnitude.  In parallel to the development of sophisticated numerical algorithms~\cite{allen2009forward,ragone2018computation} to compute rare events statistics in complex systems, here
We obtain explicit asymptotic expressions of the mean FPT to a rarely visited target, in excellent agreement with simulations. Our analysis reveals that memory effects, which so far have been left aside for this situation, can modify the kinetics by more than one order of magnitude, and finally    provides a refined characterization  of the dynamics of visits to rare configurations  for generic stationary non-Markovian processes.
  
\begin{figure}%
\includegraphics[width=6cm]{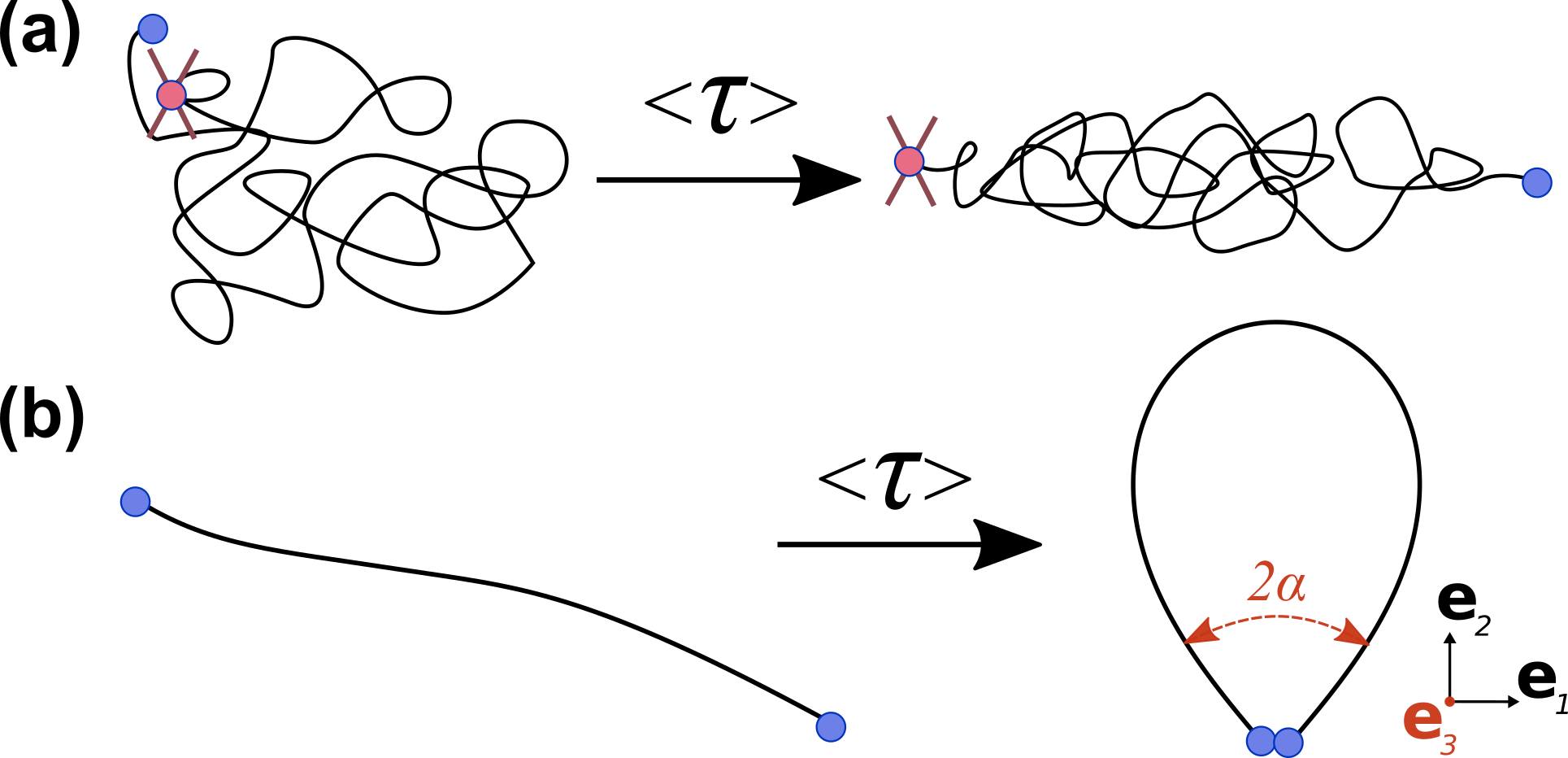}
\caption{ What is the mean first time to reach a rare configuration ? This Letter investigates this question in the case of (a) an attached flexible polymer, for which we compute the time that a large extension is reached, and (b) a stiff wormlike chain, for which we compute the time that the extremities get into contact. }
\label{FigRareEVT}%
\end{figure}

 We illustrate  our methodology by solving two problems involving polymer chains (Fig.~\ref{FigRareEVT}), which provide prototypical examples of physical  systems with many interacting  degrees of freedom, where reaction coordinates thus display memory effects \cite{Panja2010,bullerjahn2011monomer,DoiEdwardsBook}. We  calculate the mean time for a flexible chain to spontaneously reach a large extension, which is relevant in ligand adhesion via flexible tethers~\cite{jeppesen2001impact} and for the rheology of entangled melts~\cite{milner1998reptation,milner1997parameter,cao2015large}. 
We also investigate the closure kinetics of a stiff wormlike chain (i.e. a fluctuating thin rod). While this problem is highly relevant in the context of DNA looping kinetics~\cite{Vafabakhsh2012}, it has so far seemed analytically intractable, notably due to the difficulty to describe the  non-Gaussian stochastic dynamics of highly curved rods.  Existing theories for this problem  rely either on mean field approximations~\cite{ Dua2002a,guerin2014semiflexible} or on a mapping to 1-dimensional dynamics ~\cite{jun2003diffusion,Hyeon2006,Chen2004,le2013measuring}, 
which disagree with numerical simulations~\cite{Afra2013}. 

\textit{First passage for an attached flexible polymer.-} 
We first consider the simplest model of flexible polymer, formed by $N$ phantom beads, with friction coefficient $\gamma$, linked by springs of stiffness $k$. The overdamped evolution of the beads' positions $x_i$ ($i$ %\in\{1,N\}$ 
is the bead index) follows from force balance \cite{DoiEdwardsBook}
\begin{align}
\gamma\dot x_i=-k(x_{i+1}-2x_i+x_{i-1})+f_i(t) \label{EqRouse}
\end{align}  
where thermal forces obey  $\langle f_i(t)f_j(t')\rangle=2k_BT\gamma \delta(t-t') \delta_{ij}$. 
We denote by $l_0=\sqrt{k_BT/k}$ the typical bond length, and $\tau_0=\gamma/k$ the typical relaxation time of a single bond.
The first monomer is fixed, $x_1=0$, and we study the mean time $\langle\tau\rangle$ that the other polymer end $r(t)=x_N(t)$ reaches a threshold value $z$ [Fig.~\ref{FigRareEVT}(a)]. 
%This time plays a key role in the constraint release mechanism that determines the rheological properties of untangled star polymers ~\cite{milner1997parameter,milner1998reptation}, and is also involved in ligand adhesion kinetics when mediated by flexible linkers~\cite{jeppesen2001impact}. 
The energy at fixed $z$  is given by $ U=k z^2/(2N)$, and we assume $U\gg k_BT$,  so that first-passage  events to $z$ are rare.

Figure \ref{figCompCao}(a) shows  the mean FPT obtained from simulations results of Ref.~\cite{cao2015large} and  existing analytical approximations  for a fixed and relatively high value of the energy cost $ U\simeq18 k_BT$. Substantial disagreement that increases with $N$ is found, be it for   adiabatic approximations \cite{WILEMSKI1974a,cao2015large}, effective one dimensional descriptions \cite{milner1998reptation,milner1997parameter} and even the rigorous weak noise approach ($T\to 0$ at fixed $N$, see Refs.~\cite{cao2015large,schuss2009theory} and SI \cite{RefToSI}). 
This shows the necessity to take into account the collective dynamics of all monomers to calculate the  mean FPT, which is the main purpose of this work.  In fact,  the non-Markovian theory that we introduce in this paper shows an excellent agreement with simulations [Fig.~\ref{figCompCao}(a)], which holds for a broad range of values of the energy barrier [Fig.~\ref{figCompCao}(b)].

  \begin{figure}%
\includegraphics[width=8cm]{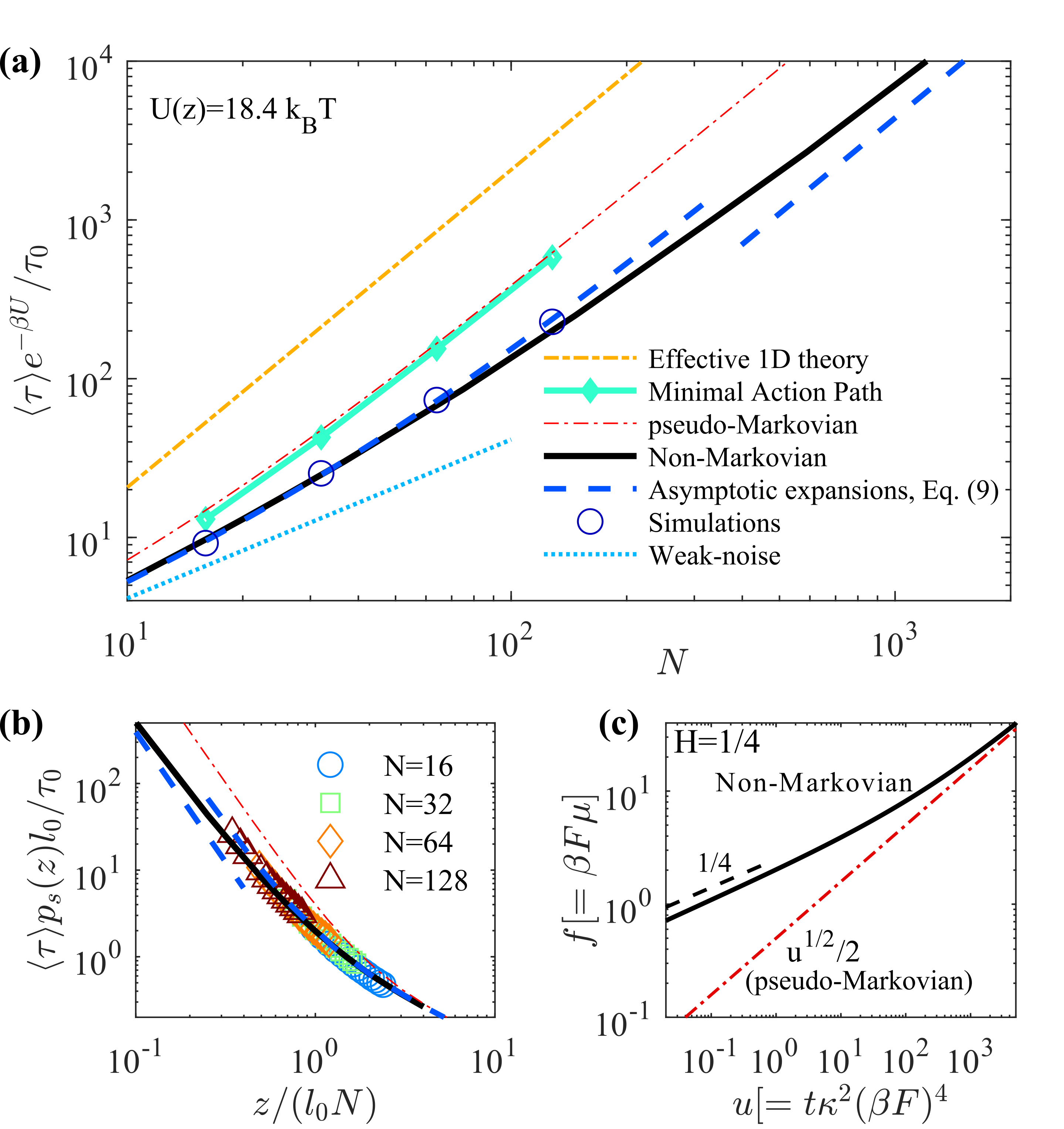}
\caption{(a) Mean FPT for a flexible chain to reach an extension $z= 3.5 l_0\sqrt{3N}$, corresponding to a fixed energy cost $U=18.4k_BT$. Symbols: simulations of Ref.~\cite{cao2015large}. Different curves correspond to different theories, obtained (from top to bottom) via a mapping over 1D dynamics (Milner-McLeish reptation theory~\cite{milner1998reptation,milner1997parameter}, upper dashed line), the Minimal Action Path method~\cite{cao2015large}, the pseudo-Markovian (Wilemski-Fixman \cite{WILEMSKI1974a}) approximation, the non-Markovian theory (this work, black thick line), asymptotic expansions of the Non-Markovian theory  [dashed blue line, Eq.~(\ref{ResultNM}), this work], and the weak-noise result $T\to0$, fixed $N$ \cite{cao2015large,schuss2009theory}. Details on all theories can be found in SI~\cite{RefToSI}. 
(b) Mean FPT in rescaled variables, with supplementary simulation data of Ref.~\cite{cao2015large} (symbols). Lines share the same color code as in (a). (c) Rescaled average trajectory $\mu(t)$ in the future of the FPT  for a scale invariant process with $H=1/4$. The dashed red line would be the future trajectory by assuming equilibrium at initial time.}
\label{figCompCao}%
\end{figure}

\textit{General expressions for the mean FPT. - } We now consider the more general problem of the FPT of a stochastic (one-dimensional) variable $r(t)$ to a rarely visited threshold $z$. We assume that $r(t)$ is non-smooth \cite{ReviewBray}, meaning that  $\langle \dot{r}^2\rangle=\infty$, as is the case for  overdamped processes. We denote $p(r,t)$ the probability density distribution of $r$ at time $t$, starting from a given initial position $r_0$ that will be proved to be irrelevant. We also assume that $r(t)$ is stationary at long times, $p(r,t)\underset{t\to\infty}\to p_s(r)$, where the stationary distribution $p_s(r)$ is reached after a finite correlation time $t_c$. With these hypotheses, the following exact expression  can be obtained~\cite{guerin2016mean}:
\begin{align}
 \langle \tau \rangle p_s(z)=\int_0^\infty dt\ [p_\pi(z,t)-p(z,t)], \label{Eq2}
\end{align}
 where  $p_\pi(r,t)$ the probability density of $r$ at a time $t$ \textit{after the first passage}. Now, in the case of targets that are only rarely visited, we stress the following key points:  
 (i) as long as $r_0$ is not in the close vicinity of $z$, $p(z,t)$ is  exponentially small (with noise intensity) at all times, and (ii) the probability $p_\pi(z,t)$ to revisit the target after a time $t$ is exponentially small at long times, but finite at times that immediately follow a FPT event, when $r$ is still close to $z$. The integral (\ref{Eq2}) is dominated by this short time contribution, where $p_\pi$ can be replaced by its value $p_\pi^\infty(z,t)$ obtained by considering the linearized dynamics around the target point. Hence, the mean FPT to a rare configuration is asymptotically (rare event limit) given by
\begin{align}
 \langle \tau \rangle p_s(z)\simeq\int_0^\infty dt \ p_\pi^\infty(z,t) \label{EqT}.
\end{align} 
Since $p_\pi^\infty(z,t)$ is a return probability for a particle submitted to a constant force in infinite space, it vanishes fast enough at long times so that the expression (\ref{EqT}) is defined without any ambiguity. 
Note that for rare events the initial distribution of $r$ has typically been forgotten long before the FPT, and thus does not influence $ \langle \tau \rangle$.  
The above equation suggests a two-step strategy to obtain $\langle \tau \rangle$. The first step consists in characterizing the static quantity $p_s(z)$ ; for equilibrium systems one obtains  $p_s(z)\propto e^{-U(z)/k_BT}$ and in particular  $\langle \tau\rangle$ follows an Ahrrenius-like law~\cite{Note1}.
The second step consists in analyzing the dynamics of $r(t)$ in the vicinity of the target  $z$ to deduce $p_\pi^\infty(z,t)$.

To proceed further, we assume that  the dynamics of $r(t)$ near $z$ is Gaussian, which is valid  in the vicinity of the most probable configuration. We denote by $m_s(t)$ and $\psi(t)$, respectively, the mean and the variance of  $r(0)-r(t)$ when the initial state is the stationary distribution conditional to $r(0)=z$.
 We adapt the theory of Ref.~\cite{guerin2016mean} (restricted to unbiased dynamics), based on the hypothesis that the trajectories followed by the random walker in the future of the FPT display Gaussian statistics. Defining the average future trajectory as $\langle r(t+\mathrm{FPT})\rangle=z-\mu(t)$ and approximating the variance in the future of the FPT by $\psi(t)$, we can write  the so-far unknown quantity $p_\pi^\infty(z,\tau)$  as 
\begin{align}\label{ppi}
p_\pi^\infty(z,t)=[2\pi\psi(t)]^{-1/2} \ e^{-\mu(t)^2/2\psi(t)}.  
\end{align}
The average future trajectory $\mu(t)$ itself satisfies the self-consistent integral equation (see SI~\cite{RefToSI})
\begin{align}
 \int_0^\infty  dt \ \frac{e^{- \mu(t)^2/(2\psi(t))}}{\psi(t)^{1/2}} \Bigg\{\mu(t+\tau)   -\mu(t)\frac{\psi(t+\tau)+\psi(t)-\psi(\tau)}{2\psi(t)}-m_s(\tau)\Bigg\}=0. \label{EqMuG1}
\end{align}
We note that our theory holds for general non-equilibrium systems. Here we focus on equilibrium ones, in which case the fluctuation-dissipation theorem imposes 
\begin{align}
m_s(t)=-\frac{  F  \psi(t)}{2k_BT}
\end{align}
where $F=-\partial_z U(z)$.  
For the Markovian (diffusive) case with $\psi(t)\propto t$, there is an obvious solution $\mu(t)= m_s(t)$. For non-Markovian variables, this relation does not hold and the  future trajectory $\mu(t)$ reflects the state of the non-reactive degrees of freedom at the FPT.  Finally, Eqs. (\ref{EqT}),(\ref{ppi}),(\ref{EqMuG1}) fully define the mean FPT to a rare configuration for general non Markovian processes that are locally Gaussian.

 In the case of a biased anomalous  dynamics with $\psi(t)=\kappa t^{2H}$, where $0<H<1$,  Eq.~(\ref{EqMuG1}) predicts that $\mu$ takes the scaling form
\begin{align}
\mu(t)=-\frac{k_BT}{F}f\left( t \ \left(\frac{\sqrt{\kappa}\  \vert F\vert}{k_BT} \right)^{1/H}\right),
\end{align}
and the mean FPT reads
 \begin{align}
 \langle \tau\rangle   p_s(z) = \frac{A_H (k_BT)^{\frac{1-H}{H}}}{  \vert F\vert ^{  \frac{1-H}{H}} \kappa^{\frac{1}{2H}}}, \ A_H=\int_0^{\infty}du \ \frac{e^{-\frac{f^2(u)}{2u^{2H}}}}{\sqrt{2\pi} u^{H}}. \label{ResultScaling}
\end{align}  
This formula provides an explicit asymptotic relation for the mean FPT, as a function of the subdiffusion coefficient $\kappa$, the local force $F$ and the  temperature $k_BT$, and $A_H$  depends only on $H$ ($f$ is defined in SI \cite{RefToSI}). Of note, this result (\ref{ResultScaling}) is consistent with the scaling proposed in Ref.~\cite{pickands1969asymptotic}  for processes that are Gaussian (not only locally). In addition, it agrees with the more recent derivation of the prefactor for this scaling based on a perturbative scheme~\cite{delorme2017pickands}  in $\varepsilon \equiv H-1/2 $ (see SI \cite{RefToSI}). We now discuss applications of these general results. 
  
\textit{Application to the kinetics of large extension for a flexible chain.-} Let us come back to the above example of an attached flexible chain. 
It is  well known that the dynamics of the ends is either diffusive, $\psi(t)=2D_0t$ for $t\ll \tau_0$, or subdiffusive, $\psi(t)=\kappa t^{1/2}$ with $\kappa=4 k_BT/(\pi \gamma  k)^{1/2}$ when $\tau_0\ll t\ll t_c$, where $t_c=N^2\tau_0$ is the correlation time. The mean FPT is controlled either by the short time diffusive regime ($H=1/2)$ or by the intermediate subdiffusive regime ($H=1/4$), so that
\begin{align}
\langle \tau\rangle p_s(z) \simeq 
\begin{cases}
0.39  (N/l_0z)^3 & (l_0\sqrt{N}\ll z\ll N l_0)\\
\frac{N}{l_0z}\left[1+ \left(\frac{N}{l_0z}\right)^2 \right] & (N l_0\ll z ) 
\end{cases}
\label{ResultNM},
\end{align}
where we have  included the (asymptotically exact) next-to-leading order expansion  in the large $z$ limit (which coincides with the weakly non-Markovian limit, see SI \cite{RefToSI}). This expression incorporates non-Markovian effects that were  neglected in Ref.~\cite{cao2015large}. Here we have used the value $A_{1/4}=2.0$, which we obtained by numerically solving Eq.~(\ref{EqMuG1}). This value is about 8 times smaller than in the pseudo-Markovian approximation (where $\mu\simeq m_s$, leading to with $A_{1/4}^{\mathrm{WF}}=16$). 
Here, the memory effects are nearly of one order of magnitude for the mean FPT and are thus strong. This  originates from the qualitative difference between the short time behaviour of the trajectory after the first passage $\mu(t)\sim t^{1/4}$ and that of $m_s(t)\sim t^{1/2}$ (following stationary state with $z=r$) [Fig.~\ref{figCompCao}(c)]. At short times $\mu(t)$ can therefore be \textit{infinitely larger} than $m_s(t)$, which means that local equilibrium assumptions are inaccurate in this situation. 
All data of the mean FPT can be collapsed on a single master curve depending only on $z/l_0N$, with asymptotics  given by Eq.~(\ref{ResultNM}). This is done in Fig.~\ref{figCompCao}(b), where we see that the simulation data closely follow (but are slightly larger than) our theoretical predictions.  
Finally, our theory provides an accurate description of the kinetics with which a flexible polymer reaches a large   extension.

\textit{The closure time of a stiff wormlike chain.} We now  consider a thin inextensible elastic rod with bending rigidity $\kappa_b$. In the stiff limit, where the persistence length $l_p=\kappa_b/k_BT$ is much larger than the contour length $L$, closure events are rare since they require overcoming a large bending energy barrier. Here we calculate  the closure time $\langle\tau\rangle$ defined as the mean time for the end-to-end distance $r$ to reach  a value $a\ll L$. We assume the dynamics to be described by the resistive force theory, in which viscous forces apply locally on the filament with friction coefficients per unit length $\zeta_\perp, \zeta_\parallel$ (respectively in the parallel and perpendicular directions) \cite{powers2010dynamics,hallatschek2007tension}. We furthermore assume that no force and no torque are exerted at the chain ends.

\begin{figure}%
\includegraphics[width=8cm]{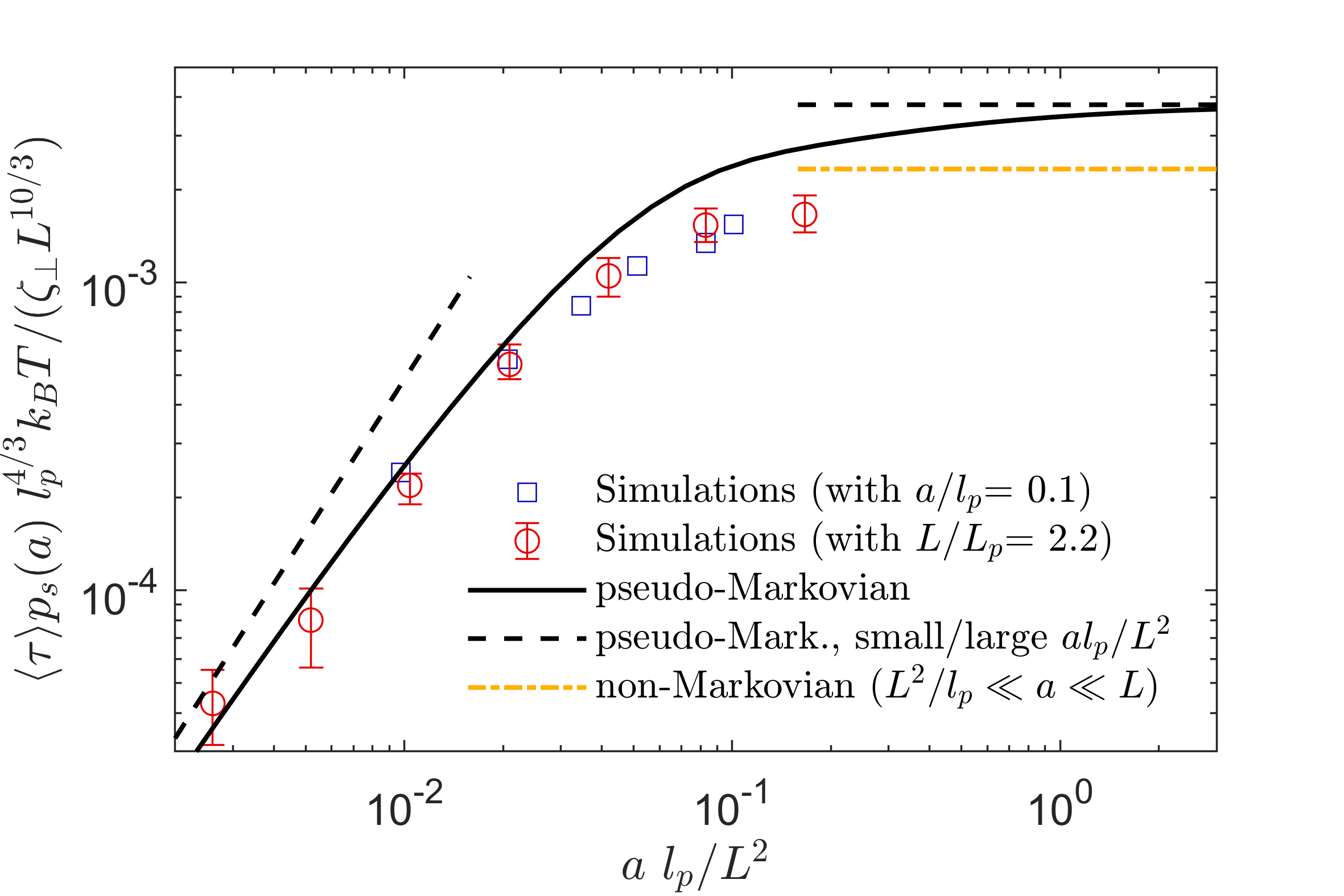}
\caption{Mean closure time for wormlike chains as a function of capture radius $a$, shown in rescaled variables. Symbols:  simulations of Ref.~\cite{Afra2013}, rescaled by $p_s(r)$ given in Ref.~\cite{mehraeen2008end}.  
Continuous black line: pseudo-Markovian approximation [Eq. (\ref{RescaleTauWLC})], with asymptotic regimes indicated by the dashed black lines. Dash-dotted orange line: non-Markovian theory for $L^2/l_p\ll a\ll L$, Eq.~(\ref{WormlikeNMLargea}). } 
\label{figWLC}%
\end{figure}

 Determining the closure time [Eq.~(\ref{EqT})] first requires to calculate $p_s(r)$, which is an equilibrium (static) statistical mechanics problem which has been studied at length by a variety of analytical and numerical methods~\cite{shimada1984,douarche2005protein,becker2010radial,guerin2017analytical,mehraeen2008end}. It is also needed to characterize the dynamics at the early times following a closure event. Such dynamics necessarily occurs at the vicinity of the close configurations of minimal bending energy. Of note,  lateral fluctuations are of the order of  $\ell_\perp(t) \propto t^{1/4}$  \cite{everaers1999dynamic,hallatschek2007tension} which is small at short times. This key remark implies that the essential of the dynamics  after closure takes place  near the extremities, where the chain can be considered as close to a straight rod. % (because of the boundary condition $\ve[r]''(0)=0$). 
We can then  calculate analytically the evolution of the end-to-end vector when initial conditions are closed equilibrium configurations. Characterizing this dynamics  in the reference frame $\{\ve[e]_i\}$ defined by the configuration at closure [see Fig.\ref{FigRareEVT}(b)] as $\ve[r]_{\mathrm{ee}}(t)=\ve[r]_{\mathrm{ee}}(0)+\sum_{i=1}^3 X_i(t)\ve[e]_i$, we obtain (see SI \cite{RefToSI}):
\begin{align}
& \langle X_1(t) \rangle =\frac{ F \ \kappa \ t^{3/4} \cos^2\alpha    }{2k_BT} ,\   \kappa =\frac{4\sqrt{2}\ k_BT}{\Gamma(7/4)\zeta_\perp^{3/4}\kappa_b^{1/4}},\nonumber \\
 &\mathrm{Cov}(X_i(t),X_j(t))=
\begin{pmatrix}
\cos^2\alpha & 0 & 0\\
 0& \sin^2\alpha & 0\\
 0 & 0 & 1 
\end{pmatrix}\ \kappa \ t^{3/4},   \label{4224}
\end{align}
where $\alpha$ is half the opening angle of the most probable closed configurations (Fig.\ref{FigRareEVT}), and the force is  $F=21.55\kappa_b/L^2=-U'(0)$, with $U(a)$ the energy cost to form a closed configuration. The stationary dynamics around a closed configuration is thus a three-dimensional biased anisotropic subdiffusion. Note that $\langle X_1(t) \rangle $ and $\mathrm{Var}(X_1)$ are again linked by the ratio $ F/2k_BT$, which is consistent with the fluctuation-dissipation theorem. A first estimate of the closure time can be obtained by assuming $p_\pi(t)\simeq p(a,t\vert a,0)$ (pseudo Markov approximation). 
This can be readily calculated from the Gaussian dynamics specified by Eq.~(\ref{4224}), leading to
\begin{align}
\langle \tau\rangle p_s(a)= \frac{\zeta_\perp \ L^{10/3}}{k_BT \  l_p^{4/3}}\Phi\left(\frac{a\ l_p}{L^2}\right), \label{RescaleTauWLC}
\end{align}
where $\Phi$ is a  scaling function calculated in SI~\cite{RefToSI} represented on Fig.~\ref{figWLC} (black line). 
This figure also displays the simulation data of Ref.~\cite{Afra2013}, which collapse as in Eq.~(\ref{RescaleTauWLC}) onto a curve which is close to $\Phi$ for small arguments. We stress that there is no fitting parameter in the theory. 
 
However, there is a difference of a factor of about two between theory and numerics for larger capture radius, suggesting that non-Markovian effects are significant in the regime  $a\gg L^2/l_p$, which we investigate now 
(while still keeping the small capture radius condition $a\ll L$). In this case the dynamics needs to be characterized only at time scales where the return probability is not exponentially small, i.e. such that $\langle X_1(t) \rangle^2$ is smaller than $\langle X_1^2(t) \rangle$. For these time scales, $X_1\sim L^2/l_p$ is still much smaller than $a$. This implies that the end-to-end distance is approximated at linear order as $r=[(a+X_1)^2+X_2^2+X_3^2]^{1/2}\simeq a + X_1$ and is thus equivalent to a one dimensional Gaussian variable. The mean closure time can be obtained  by applying the formalism presented above with $H=3/8$. We obtain
\begin{equation}
\langle \tau\rangle p_s(a) \simeq 0.0023 \ \frac{\zeta_{\perp} L^{10/3}}{k_BT\ l_p^{4/3} }    \hspace{0.2cm} (a\gg L^2/l_p).\label{WormlikeNMLargea}
\end{equation} 
Here the value of the prefactor was obtained with $A_{3/8}=2.1$ which is  $1.6$ times smaller than its estimate in the pseudo-Markovian (Wilemski-Fixman) approximation $A_{3/8}^{\mathrm{WF}}=3.39$. This explains why the pseudo Markovian theory overestimates the simulation data.  

In the opposite limit $a\ll L^2/l_p$, the pseudo-Markovian expression (\ref{RescaleTauWLC}) becomes 
\begin{equation}
\langle \tau\rangle p_s(a) \simeq  1.05  \frac{ \ \zeta_{\perp} a^{5/3} l_p^{1/3}}{k_BT  }  \hspace{0.5cm} (a\ll L^2/l_p)\label{RegimeSmallCaptureRadius}, 
\end{equation}
and it can be shown that this result can be found by setting $F=0$, i.e. by analyzing a symmetric anisotropic three dimensional subdiffusive walk. In a recent work~\cite{guerin2014semiflexible} for a similar (but isotropic) subdiffusive process, it was shown  that memory effects led to a slight reduction ($15\%$) of the mean FPT. We expect a similar for the mean closure time, as confirmed by the comparison with numerical simulations in Fig.\ref{figWLC}. 
   
%Note that in both cases the numerical prefactors are very different from unity and must be estimated carefully. 
    
\textit{Conclusion.-} In this Letter, we have introduced theoretical tools to determine the mean FPT to rarely visited configurations for generic non-Markovian processes. We have derived explicit asymptotic expressions for  the closure kinetics of a stiff wormlike chain, and for the mean FPT to a large extension of a flexible chain. 
 As demonstrated by the example of wormlike chain closure, the dynamics  needs to be Gaussian only locally (in the vicinity of the target) to apply our theory. This approach shows quantitatively the importance  of memory effects on mean FPTs, and thereby significantly improves  existing theories, whether based on a weak-noise limit,  a mapping on one dimensional problems or pseudo-Markovian (adiabatic) approximations. Our approach  is not limited to polymers, and can apply to generic complex physical systems, where the dynamics of a reaction coordinate is coupled to many other degrees of freedom.   
  
\acknowledgments{O.B. acknowledges the support of  the European Research Council starting Grant No. FPTOpt-277998. We thank A. Spakowitz for providing the data of Ref.~\cite{mehraeen2008end}. }

\newpage

\begin{center}
\large \textbf{Supplemental Material} 
\end{center} 
 
\,

\,

 \appendix
 
This Supplemental Material presents details of calculations supporting the main text. We provide 
\begin{itemize}
\item details on the equations of the non-Markovian theory for rare first passage times (Appendix \ref{GenTheor}),
\item details on asymptotically exact results  for weakly non-Markovian processes (Appendix \ref{SectionWeaklyNM}), 
\item details for the first passage problem of the flexible chain, and how existing results in the literature can be understood within our formalism, leading to the curves that are presented in Fig.~2 (Appendix \ref{SI_Flexible}),
\item calculation details for the problem of wormlike chain closure (Appendix \ref{SectionWLC_SI}).
\end{itemize}

\,
 
\section{Non-Markovian theory for first passage times in the rare event limit}
\label{GenTheor}
\subsection{General non-Markovian theory}
We consider a general one-dimensional stochastic variable $r(t)$ evolving in continuous time $t$. We denote $p(r,t)$ the probability density distribution of $r$ at time $t$. We assume that $r(t)$ is non-smooth (this means that $\langle \dot{r}^2\rangle=\infty$: the stochastic trajectories are not derivable, as is the case for the overdamped processes we have in mind). We also assume that $r(t)$ is stationary at long times, $p(r,t)\underset{t\to\infty}\to p_s(r)$, where the stationary distribution $p_s(r)$ is reached after a finite correlation time $t_c$. We aim to calculate the statistics of the first time that $r(t)$ reaches a threshold $z$, which can be reached only with a high energy cost.  We start by writing the general equation
\begin{align}
p(z,t)=\int_0^t \ d\tau \ p(z,t\ \vert\ \mathrm{FPT}=\tau)\ f(\tau) \label{EqR},
\end{align}     
where $f(\tau)$ is the First passage time (FPT) distribution and $p(z,t\vert \mathrm{FPT}=\tau)dr$ is the probability of observing $r\in[z,z+dr]$ at $t$ given that the FPT is $\tau$. We substract $p_s(r)$ on both sides of Eq.~(\ref{EqR}) and integrate over time from $t=0$ to $\infty$. 
\begin{align}
	\int_0^\infty dt [p(z,t)-p_s(z)]=- p_s(z)\int_t^\infty d\tau f(\tau)+\int_0^t d\tau f(\tau) [p(z,t\vert \mathrm{FPT}=\tau)-p_s(z)].\label{845931}
\end{align}
where we note that all the above integrals exist, since propagators converge exponentially fast towards $p_s$, due to the hypothesis of finite correlation time. Next, we write the following identities:
\begin{align}
\int_0^\infty dt \int_0^t d\tau f(\tau) [p(z,t\vert \mathrm{FPT}=\tau)-p_s(0)]&=\int_0^\infty d \tau \int_0^\infty du \ f(\tau) [p(z,\tau+u\vert \mathrm{FPT}=\tau)-p_s(z)]\nonumber\\
&= \int_0^\infty du \int_0^\infty d \tau \ f(\tau) [p(z,\tau+u\vert \mathrm{FPT}=\tau)-p_s(z)]\nonumber\\
&= \int_0^\infty du \left\{\int_0^\infty d \tau \ f(\tau) p(z,\tau+u\vert \mathrm{FPT}=\tau)-\int_0^\infty d \tau \ f(\tau) p_s(z)\right\}\nonumber\\
&=\int_0^\infty du\ [p_{\pi}(z,u)-p_s(z)], \nonumber 
\end{align}
where at each line all integrals are well defined because of the convergence of the propagators towards $p_s(z)$. Note that we have used the definition of $p_\pi$ as the probability density of $r$ at a time $t$ after the FPT, so that
\begin{align}
p_{\pi}(z,t)=\int_0^\infty d \tau \ f(\tau) p(z,\tau+t\vert \mathrm{FPT}=\tau)
\end{align}
 We also note that
\begin{align}
	\int_0^\infty dt\int_t^\infty d\tau f(\tau)=\int_0^\infty d\tau f(\tau) \int_0^\tau dt =\int_0^\infty d\tau f(\tau) \tau = \langle\tau\rangle
\end{align}
with $\langle\tau\rangle$ the Mean First Passage Time to the target. Eq.~(\ref{845931}) becomes  
\begin{align}
 \langle \tau \rangle p_s(z)=\int_0^\infty d\tau [p_\pi(z,\tau)-p(z,t)], \label{5042}
\end{align}
This exact relation~\cite{guerin2016mean} is the starting point of our analysis, for any non-smooth stochastic process $p(r)$ that becomes stationary at long times. 
Now, in the case of targets that are only rarely visited, we note that probability density $p_\pi(z,\tau)$ to revisit the target after a time $\tau$ is largest at the times that immediately follow the FPT events (when $r$ is still close to $z$) and becomes exponentially small (with $k_BT$) for larger times. Furthermore, if the starting point $r_0$ is not too close to the target, the probability density $p(z,t)$ to reach the target, starting from the initial state, will be exponentially small at all times. These remarks suggests that we can evaluate the integrals in Eq.~(\ref{5042}) by replacing $p_\pi$ by its value calculated by linearizing the dynamics around the most probable state, in which case we denote it as $p_\pi^\infty$, and that we can neglect the second term $p(z,t)$, leading to  
\begin{align}
 \langle \tau \rangle p_s(z)=\int_0^\infty d\tau \  {p_\pi ^\infty}(z,\tau) ,  \label{EqTAppendix}
\end{align}
which gives the mean FPT in the limit of rare events. The fact that the mean FPT does not depend on the initial distribution $p_0(r)$ is consistent with the intuition that the mean FPT to a rare configuration is in general much longer than the correlation time, so that the initial distribution has typically been forgot long before the FPT. 
Note that when the dynamics is linearized around $z$, $r(t)$ is equivalent to a particle submitted to a constant force in infinite space, so that $p_\pi^\infty(z,t)$ vanishes for large times, and the integral  (\ref{EqTAppendix}) is  defined without ambiguity. 
The rare event limit is similar to the large volume approximation for symmetric (non-compact) random walks  in confinement~\cite{Condamin2007,guerin2016mean}. Physically, one should include an upper cutoff in the integral (\ref{EqTAppendix}) of the order of the correlation time, where the approximation of linearized dynamics does not hold anymore. However, since at such times $p_\pi$ is already exponentially small there is no need to introduce explicitly this cutoff.  Note also that, at long times, both terms $p_\pi(z,t)$ and $p(z,t)$ approach the stationary probability $p_s(z)$ and compensate each other, which is a supplementary argument to take only short times into account in Eq.~(\ref{EqTAppendix}).   
 
This fact can be illustrated by considering the  simple case that $r(t)$ is a one-dimensional Brownian motion, with diffusion coefficient $D$ and friction coefficient $\gamma=k_BT/D$ in a potential $U(r)$. Since this process is memoryless, $p_\pi^\infty(t)$ is directly given by the propagator $p^\infty(z,t\vert z,0)$  calculated by considering the linearized dynamics around $z$, for which
\begin{align}
p_\pi^\infty(t)=p^\infty(z,t\vert z,0)=\frac{e^{-(F t /\gamma)^2/[4 D t]}}{[4\pi D t]^{1/2}},
\end{align} 
which presents an exponentially fast decay when $t\to\infty$. Here, $F=-\partial_r U\vert_{r=z}$ is the local potential slope. Inserting this value into Eq.~(\ref{EqTAppendix}) leads to
\begin{align}
 \langle \tau \rangle p_s(z)\simeq \frac{\gamma}{\vert F\vert}\hspace{1cm}\text{(1D diffusive particle in a potential)} \label{1DRes}
\end{align}
This expression is consistent with existing results for this Markovian case which was early studied by Kramers~ \cite{Kramers1940}, see Ref.~\cite{hanggi1990reaction}.

Coming back to the non-Markovian case, we are left to evaluate the probability to return to the target at a time $t$ after the FPT. To characterize it, we adopt the strategy of Ref.~\cite{guerin2016mean} where it was shown that memory effects can be taken into account by characterizing the trajectories in the future of the FPT. %Let us define $\mu(t)$ so that $\langle r(t+\mathrm{FPT})\rangle=z-\mu(t)$. 
We write the generalization of (\ref{EqR}) for 2 points, for any fixed $t_1>0$, 
\begin{align}
p(z,t;r_1,t+t_1)=\int_0^t \ d\tau \ p(z,t;r_1,t+t_1\ \vert\ \mathrm{FPT}=\tau)\ f(\tau) ,\label{8421}
\end{align}     
where $p(z,t;r_1,t+t_1)$ is the joint probability density of $r=z$ at time $t$ and $r=r_1$ at $t+t_1$ ; for large times this quantity does not depend on $t$ and thus defines the stationary probability $p_s(z,0;r_1,t_1)$ to observe $z$ at an initial time and $r_1$ after a subsequent time $t_1$. Taking the Laplace transform of Eq.~(\ref{8421}) and considering small values of the Laplace variable leads to
\begin{align}
\langle \tau \rangle p_s(z,0;r_1,t_1)=\int_0^\infty d\tau [p_\pi(z,\tau;r_1,\tau+t_1)-p_0(z,\tau;r_1,\tau+t_1)].
\end{align}
In the limit of rare events, using the same assumptions as in Eq.~(\ref{EqTAppendix}), we obtain
\begin{align}
\langle \tau \rangle p_s(z,0;r_1,t_1)=\int_0^\infty d\tau\  p_\pi^\infty(z,\tau;r_1,\tau+t_1).
\end{align}
Integrating over $r_1$ leads to
\begin{align}
\langle \tau \rangle p_s(z) m_s(t_1)=\int_0^\infty d\tau \ p_\pi^\infty(z,\tau) \mu(t_1+\tau\vert \tau) 
\end{align}
which generalizes Eq.(\ref{EqTAppendix}). Here, $\mu(\tau+t_1\vert  \tau) $ is the average of $z-r(\tau+t_1+\mathrm{FPT})$ given that $r(\tau+\mathrm{FPT})=z$.  We have also defined $m_s(t)$ so that  so that  $\langle r(t)\rangle_{r(0)=z} = z-m_s(t)$, where $\langle\cdot\cdot\cdot\rangle_{r(0)=z}$ is the equilibrium average conditional to $r(0)=z$. Note that this equation is still \textit{exact} in the rare events limit for any non-smooth stochastic process. 

Let us now call $\psi(t)$ the mean square Displacement function i.e., $\psi(t)$ is the variance of $r(t)-r(0)$ conditional to $r(0)=z$. 
An explicit equation that defines $\mu(t)$ in a self-consistent way can be found by assuming that the distribution of paths $p_\pi$ in the future of the FPT is Gaussian, and that its covariance in the future of the FPT is the stationary covariance, so that
\begin{align}
&\int_0^\infty  dt \ \frac{e^{- \mu(t)^2/(2\psi(t))}}{\psi(t)^{1/2}} \Bigg\{\mu(t+\tau)  -\mu(t)\frac{\psi(t+\tau)+\psi(t)-\psi(\tau)}{2\psi(t)}-m_s(\tau)\Bigg\}=0 \label{EqMuG}
\end{align}
It is very important to note that $\mu(t)\neq m_s(t)$ because the configurations of the monomers are not at equilibrium at the FPT instant. One exception is the case $\psi(t)\propto t$ (locally diffusive process), for which $\mu(t)=m_s(t)$ is a trivial solution of Eq.~(\ref{EqMuG}).
The mean FPT can be found from Eq.~(\ref{EqTAppendix}) in our Gaussian closure approximation as 
\begin{align}
\langle \tau \rangle p_s(z)\simeq\int_0^\infty d\tau   \frac{e^{- \mu(t)^2/(2\psi(t))}}{[2\pi\psi(t)]^{1/2}}.
\end{align}
Finally, in the case of equilibrium systems, when the dynamics of $r$ at the vicinity of $z$ is Gaussian, $m_s(t)$ is linked to the MSD function by
\begin{align}
m_s(t)=-\psi(t)\frac{F}{2k_BT}, \label{FluctDiss}
\end{align}

This can be shown by considering the description of the dynamics by a Generalized Langevin Equation where the force applied on the particle is $F=-U'(z)$,  so that
\begin{align}
\int_0^t d\tau K(t-\tau)\dot{r}(\tau)=F +g(t),  \hspace{2cm} \langle g(t)g(t')\rangle =k_BT K(\vert t-t'\vert),  \label{EqGLE}
\end{align}
where $\dot{r}(t)$ is the velocity, $K$ a memory (friction) kernel and we have neglected inertia. Assuming the starting position to be equal to zero, this equation becomes in Laplace space
\begin{align}
s \tilde{K}(s) \tilde{r}(s)=F/s +\tilde{g}(s) 
\end{align}
so that the average trajectory reads (in Laplace space)
 \begin{align}
\langle \tilde r(s)\rangle= F [s^2 \tilde{K}(s)]^{-1} \label{AvTraj}
\end{align}
In turn, the covariance of the trajectories reads (in Laplace space)
\begin{align}
\tilde{\sigma}(s,s')=\int_0^\infty dt\int_0^\infty dt' e^{-st-s't'}[\langle  {r}(t) {r}(t')\rangle-\langle  {r}(t)\rangle \langle {r}(t')\rangle]=
\langle \tilde{r}(s)\tilde{r}(s')\rangle-\langle \tilde{r}(s)\rangle \langle\tilde{r}(s')\rangle = \frac{\langle \tilde{g}(s) \tilde{g}(s')\rangle}{ss'\tilde{K}(s)\tilde{K}(s')} 
\end{align}  
Using Eq.~(\ref{EqGLE}), we find
\begin{align}
\tilde{\sigma}(s,s')=\frac{k_BT}{ss'\tilde{K}(s)\tilde{K}(s')} \int_0^\infty dt \int_0^\infty dt' e^{-st-s't'}K(\vert t-t'\vert)
= k_BT\frac{\tilde{K}(s)+\tilde{K}(s')}{(s+s')ss'\tilde{K}(s)\tilde{K}(s')} \label{Sigma0}
\end{align}  
Let us consider $\sigma(t,t')$ of the form
\begin{align}
\sigma(t,t')=\frac{1}{2}[\psi(t)+\psi(t')-\psi(\vert t-t'\vert)]
\end{align}
for which the double Laplace transform is
\begin{align}
\tilde{\sigma}(s,s')=\int_0^\infty dt \int_0^\infty dt' e^{-st-s't'} \sigma(t,t')=\frac{1}{2}\left[\frac{\tilde{\psi}(s)}{s'}+\frac{\tilde{\psi}(s')}{s}-\frac{\tilde{\psi}(s)+\tilde{\psi}(s')}{s+s'}\right]=\frac{1}{2}\left[\frac{\tilde{\psi}(s)s^2+\tilde{\psi}(s')(s')^2}{s's(s+s')}\right]\label{SigmaPsi}
\end{align}
Comparing Eqs.~(\ref{Sigma0}), (\ref{SigmaPsi}) and (\ref{AvTraj}) leads to
\begin{align}
\tilde{\psi}(s)=\frac{k_BT}{2\tilde{K}(s)s^2}= \frac{k_BT }{2F}\langle \tilde r(s)\rangle 
\end{align}
The above equation is the Laplace transform of Eq. (\ref{FluctDiss}), which is thus a consequence of the fluctuation dissipation theorem.   
 
\subsection{Scale invariant processes and Pickands' constants}

Let us now illustrate our theory in the case where the local dynamics is locally a biaised anomalous diffusion, with $\psi(t)=\kappa t^{2H}$, $H$ being the Hurst exponent, $0<H<1$. In this case, Eq. (\ref{EqMuG}) predicts that $\mu$ takes the scaling form
\begin{align}
\mu(t)=\frac{1}{\beta F}f\left( t \ (\sqrt{\kappa} \beta \vert F\vert )^{1/H}\right)
\end{align}
where $f$ is solution of 
\begin{align}
\int_0^\infty dt &\ \frac{e^{- f(t)^2/(2t^{2H})}}{t^{H}} \Bigg\{f(t+\tau)  -f(t)\frac{(t+\tau)^{2H}+t^{2H}-\tau^{2H}}{2t^{2H}} -\frac{\tau^{2H}}{2}\Bigg\}=0 \label{EqMuH}
\end{align}
and the mean FPT reads
\begin{align}
&\langle \tau\rangle   p_s(z) = \frac{1}{(\beta \vert F\vert)^{  \frac{1-H}{H}} \kappa^{\frac{1}{2H}}}A_H \label{T_AH} \\
&A_H=\int_0^{\infty}dt \ e^{-f^2(t)/2t^{2H}}/(2\pi t^{2H})^{1/2} \label{AH}
\end{align}  
This formula provides an explicit asymptotic relation for the mean FPT, as a function of the subdiffusion coefficient $\kappa$, the local force $F$ and the inverse temperature $ \beta$. $A_H$ is a constant that only depends on $H$. 
 
In the mathematical literature~\cite{pickands1969asymptotic}, the case that $r(t)$ is a one dimensional Gaussian process at all times (not only at the vicinity of the target) has been analyzed in the rare event limit. We find that for this subclass of processes our expression (\ref{T_AH}) is compatible with the analysis of Ref.~\cite{pickands1969asymptotic} if we identify
\begin{align}
A_H=2^{1/(2H)}/\mathcal{H}_{2H}
\end{align}
where the so-called Pickands' constants $\mathcal{H}_{\alpha}$  depend only on $\alpha=2H$. Our theory suggests that Pickands' constants could be used to characterize the FPT kinetics of non-Gaussian processes. Our theory also provides approximates expressions for those constants, which are difficult to estimate numerically.  Here we note that our theory reproduces exactly the recent exact perturbative results of Delorme et al.~\cite{delorme2017pickands} (see next Section).
  
\section{Perturbation theory for weakly non-Markovian processes}
\label{SectionWeaklyNM}
  
  \subsection{Exactness of the theory at first order}
Here we give an argument suggesting that our theory is exact at first non-trivial order for weakly non-Markovian processes. We consider the generalization of Eq. to any path $y(\tau)$ starting at $y(0)=0$, for which the following equation is exact in the rare event limit 
\begin{equation}
	\langle\tau\rangle p_s(0) P_{\text{stat}}([y(\tau)]\vert y(0)=0)=\int_0^{\infty}dt  \ \Pi([y(\tau)],t)  \label{Eq989},
\end{equation}
where the target position is here $z=0$, $P_{\text{stat}}([y(\tau)]\vert y(0)=0)$ is the stationary probability to follow the path $[y(\tau)]$ conditional to $y(0)=0$, $\Pi([y(\tau)],t)$ is the probability to follow the path $[y]$ in the future $t$ of the FPT (ie, the probability that $x(FPT+\tau+t)=y(\tau)$ for all $\tau>0$). Let us define the functional
\begin{align}
&	\mathcal{F}([k])=
\int\mathcal{D}[y]e^{i\int_0^\infty d\tau k(\tau)y(\tau)} \left\{\langle\tau\rangle p_s(0) P_{\text{stat}}([y(\tau)]\vert y(0)=0)-\int_0^{\infty}dt\   \Pi([y(\tau)],t)   \right\},
\end{align}
which should vanish for all $[k(\tau)]$. Let us evaluate this functional for the case that $\Pi$ is a Gaussian distribution, of mean $\mu(t)$ and covariance $\gamma(t,t')$. In this case, using formulas of Gaussian integration, we get
\begin{align}
\mathcal{F}([k(\tau)])= \int_0^{\infty}dt  \ 
\frac{e^{-[\mu(t) ]^2/2\gamma(t,t)}}{[2\pi\gamma(t,t)]^{1/2}}  \Big[ &e^{i \int_0^{\infty}d\tau k(\tau)A_{\pi}(t,\tau)-\frac{1}{2}\int_0^{\infty}d\tau\int_0^{\infty}d\tau'k(\tau)k(\tau')B_{\pi}(t,\tau,\tau')}\nonumber\\
&-e^{i \int_0^{\infty}d\tau k(\tau)m_s(\tau)
-\frac{1}{2}\int_0^{\infty}d\tau\int_0^{\infty}d\tau'k(\tau)k(\tau')\sigma_s(\tau,\tau')}
\Big].\label{89100}
\end{align}
Here the stationary paths contional to $y(0)=0$ are assumed to have a mean $m_s(t)$ and covariance $\sigma_s(t,t')$, and the quantities $A_\pi$ and $B_\pi$ are defined as 
\begin{align}
A_{\pi}(t,\tau) =\mu(t+\tau)-\mu(t)\frac{\gamma(t+\tau,t)}{\gamma(t,t)}, \hspace{1cm}  
B_{\pi}(t,\tau,\tau') =\gamma(t+\tau,t+\tau')-\frac{\gamma(t+\tau,t)\gamma(t+\tau',t)}{\gamma(t,t)}.\label{ExpressionBPi} 
\end{align}
Following an equilibrium condition, the stationary covariance and the mean satisfies the property
\begin{align} 	 
	\sigma_s(\tau,\tau')=[\psi(\tau)+\psi(\tau')-\psi(\vert \tau-\tau'\vert)]/2  ,\hspace{0.5cm}
	m_s(\tau)=\vert F\vert \psi(\tau)/2k_BT
\end{align}
Consider the case of weakly non-Markovian processes, for which
\begin{align}
\psi(t)=t+\varepsilon \psi_1(t)+\mathcal{O}(\varepsilon^2)
\end{align}
where $\varepsilon\to0$ is a small parameter, $\psi_1(t)$ is the deviation of the MSD at first order with respect to the diffusive (Markovian) case, and we have chosen our units so that the diffusion coefficient for $\varepsilon=0$ is equal to $1/2$. We also set the length scale so that $k_BT/F=1$. 

We now show that the functional $\mathcal{F}([k(\tau)])$ vanishes for all functions $[k(\tau)]$ at first order in $\varepsilon$ if $\mu(t)$ and $\gamma(t,t')$ admit the expansion 
\begin{align}
\mu(t)=t/2+\varepsilon [\psi_1(t)/2+\mu_1(t)]+\mathcal{O}(\varepsilon^2),
\gamma(t,t')= \frac{t+t'-\vert t-t'\vert}{2}+\varepsilon \gamma_1(t,t')+\mathcal{O}(\varepsilon^2)
\end{align}
with appropriately chosen functions $\mu_1(t)$ and $\gamma_1(t,t')$. Note that $\mu_1(t)$ is the deviation at first order with respect to $m_s(t)$. Indeed, the functional  $\mathcal{F}([k(\tau)])$ vanishes at order $\varepsilon^0$, and its expansion at order $\varepsilon$ reads: 
\begin{align}
& \mathcal{F}([k ])=  \varepsilon \int_0^{\infty}dt   \ 
\frac{e^{-t/8}}{(2\pi t)^{1/2}}\Big\{
 i \int_0^{\infty}d\tau k(\tau)\left[ \mu_1(t+\tau)-\mu_1(t)+\frac{\psi_1(t+\tau)-\psi_1(t)}{2}+\frac{\gamma_1(t+\tau,t)-\gamma_1(t,t)}{2t}-\frac{\psi_1(\tau)}{2}\right]\nonumber\\
 &-\iint_0^{\infty}d\tau d\tau' \frac{k(\tau)k(\tau')}{2} \left[ \gamma_1(t+\tau,t+\tau')-\gamma_1(t,t+\tau')-\gamma_1(t+\tau,t)+\gamma_1(t,t)-\frac{\psi_1(\tau)+\psi_1(\tau')-\psi_1(\vert \tau-\tau'\vert)}{2}\right]
  \Big\}\nonumber\\
  &+\mathcal{O}(\varepsilon^2)\label{kgre}
\end{align}
 We remark that the quadratic terms in $k$ vanish if one imposes the covariance of paths in the future of the FPT to be equal to the stationary covariance,
 \begin{align}
 \gamma_1(t,t')=\frac{\psi_1(\tau)+\psi_1(\tau')-\psi_1(\vert \tau-\tau'\vert)}{2}.
 \end{align}
If we use this result, we see that the linear term in $k$ in Eq.~(\ref{kgre}) also vanishes if $\mu_1$ satisfies the integral equation
 \begin{align}
\int_0^\infty dt \ K(t) \left[\mu_1(t+\tau)-\mu_1(t)\right] = 
G(\tau) \label{IntegralEq}
\end{align}
with 
\begin{align}
K(t)=\frac{e^{-t/8}}{\sqrt{t}}, \hspace{1cm} G(\tau)=- \frac{1}{4}\int_0^\infty dt \frac{e^{-t/8}}{\sqrt{t}}\left[\psi_1(t+\tau)-\psi_1(t)-\psi_1( \tau)\right]. \label{DefG}
\end{align}
This means that our theory becomes exact at order $\varepsilon$ if we impose $\mu_1$ to be the solution of Eq.~(\ref{IntegralEq}). 

\subsection{Explicit expressions at first order}
      
We now derive the explicit solution for $\mu_1$ (and thus of the mean FPT) for weakly non-Markovian processes. We take the derivative of Eq.~(\ref{IntegralEq}) with respect to $\tau$ and obtain  
\begin{align}
\int_0^\infty dt \ K(t)\ \mu_1'(t+\tau)=G'(\tau)\equiv g(\tau).
\end{align}
The formal solution of the above equation can be written in Laplace space as \cite{HandbookIntegralEquations}
\begin{align}
\widetilde{[\mu_1']}(s)=\frac{\tilde{g}(s)}{\tilde{K}(-s)},
\end{align}
where $\tilde{K}(s)=\sqrt{\pi/(s+1/8)}$ is the Laplace transform of $K(t)$. We now use the Mellin-Bromwich formula and the definition of the Laplace transform to write
\begin{align}
\mu_1(t)=\int_{-\infty}^\infty \frac{d\omega}{2\pi} \frac{[e^{i\omega t}-1]}{i\omega \tilde{K}(-i \omega)}\int_0^\infty dx\ e^{-i\omega x} g(x),
\end{align}
where we have integrated once over $t$ and used $\mu_1(0)=0$.
 If we change the order of integration and change $\omega\to-\omega$, we realize that 
\begin{align}
\mu_1(t)=\int_0^\infty dx\  g(x) [W(x-t)\Theta(x-t)-W(x)], \label{08542}
\end{align}
where $\Theta$ is the Heaviside step  function and $W$ is the inverse Laplace transform of $\tilde{W}(s)=[s\tilde{K}(s)]^{-1} $, so that
\begin{align}
W(x)=\frac{e^{-x/8}}{\pi \sqrt{x}} + \mathrm{Erf}\left(\frac{\sqrt{x}}{2 \sqrt{2 }}\right)\frac{1}{2\sqrt{2\pi}}.
\end{align}
Noting that $W$ goes exponentially fast to a constant $W(x\to\infty)=W_\infty=1/\sqrt{8\pi}$, and making sure that we do not separate illegally the two parts in the integral (\ref{08542}), we obtain
\begin{align}
\mu_1(t)=W_\infty G(t)-\int_0^\infty dx \ [g(x+t)-g(x)][W(x)-W_\infty].
\end{align}
Using Eq.~(\ref{DefG}) and $g=G'$ leads to
\begin{align}
\mu_1(t)=W_\infty G(t)+\frac{1}{4}\int_0^\infty dx \int_0^\infty du\  [W(x)-W_\infty]\frac{e^{-u/8}}{\sqrt{u}}[\psi_1'(u+x+t)-\psi_1'(x+t)-\psi_1'(u+x)+\psi_1'(x)].
\end{align}
We now integrate by parts over $x$ (taking care of chosing primitives that vanish for $x=0$, so that integrals exist):
\begin{align}
&\mu_1(t)=  - \frac{1}{4\sqrt{8\pi}}\int_0^\infty du \frac{e^{-u/8}}{\sqrt{u}}\left[\psi_1(u+t)-\psi_1(t)-\psi_1( u)\right]\nonumber\\
&+\frac{1}{8\pi}\int_0^\infty dx \int_0^\infty du  \frac{e^{-[x+u]/8}}{x^{3/2}\sqrt{u} }[\psi_1(u+x+t)-\psi_1(u+t)-\psi_1(x+t)+\psi_1(t)-\psi_1(u+x)+\psi_1(u)+\psi_1(x)] \label{0542},
\end{align}
where we have used $W'(x)=-e^{-x/8}/[2\pi x^{3/2}]$. The reactive trajectory $\mu_1$ can thus be obtained as integrals involving the MSD $\psi_1(t)$ at first order and simple functions. 

This explicit expression of $\mu_1(t)$ can now be used to obtain an expansion of the mean FPT via the formula
\begin{align}
\langle \tau\rangle p_s(z)=\int_0^\infty dt \frac{e^{-\mu(t)^2/2\psi(t)}}{[2\pi\psi(t)]^{1/2}}=2-\varepsilon \int_0^{\infty}dt \frac{e^{-t/8}}{2\sqrt{2\pi t}}\left[\mu_1(t)+\frac{\psi_1(t)}{t}+\frac{\psi_1(t)}{4}\right]+\mathcal{O}(\varepsilon^2) \label{0543}.
\end{align}

\subsection{Applications of the perturbation theory}
\textit{Application: Pickand's constants at first order.- } In the case of the baised subdiffusion, the MSD is $\psi(t)=t^{2H}$, with $H=1/2+\varepsilon$, so that the function $\psi_1$ is readily identified to be 
\begin{align}
\psi_1(t)=2 \ t\ln t. 
\end{align}
Inserting the expression  into Eq.~(\ref{0542},\ref{0543}) we see that the mean FPT can be obtained by evaluating double and triple integrals. In this case, the normalized mean FPT is called $A_H$ in the main text and reads
\begin{align}
A_H=2+4\varepsilon(\gamma_e -\ln2),
\end{align}
with $\gamma_e$ the Euler-Mascheroni constant. 
This result is compatible at first order in $\varepsilon$ with the exact result of Ref \cite{delorme2017pickands} when we identify $A_H=2^{1/(2H)}/\mathcal{H}_{2H}$. 

\textit{Large deviation of the flexible chain for very large extensions.} A second application of the perturbation theory is the large deviation of the Rouse chain in the large force limit, i.e. $F=kz/N\gg k_BT/l_0$. In this limit, only small times of the MSD function are relevant, which can be seen by defining the relevant length scale $L_s=k_BT/F$ and time scale $t_s=L_s^2/2D$, for which the rescaled MSD   (\ref{psiRouse}) becomes
\begin{align}
\tilde{\psi}(\tilde{t})=\frac{1}{L_s^2}\psi(t_s \tilde{t})\underset{F\to\infty}{\sim}\tilde{t}-\frac{\tilde{t}^2}{4F^2}+...
\end{align}
Now, using $1/F^2$ as our small parameter, and $\psi_1(t)=-t^2/4$, we obtain from Eqs.~(\ref{0542},\ref{0543})
\begin{align}
\langle \tau\rangle p_s(z)=\frac{t_s}{L_s}(2+2/F^2)
\end{align} 
so that, reestablishing homogeneity:
\begin{align}
\langle \tau\rangle p_s(z)\underset{z/[Nl_0]\to\infty}{\sim}\frac{\zeta N }{k z}\left(1+\frac{l_0^2N^2}{z^2}+...\right)
\end{align} 
and we thus obtain an explicit expression of the mean FPT which includes exact non-Markovian effects at leading order.   
 
\section{The time to reach a large extension for a Gaussian flexible chain}
\label{SI_Flexible}

\subsection{Characterization of the dynamics}

We now present calculation details for the first passage problem for the flexible phantom chain model.  We remind here the equations for the dynamics of a chain of $N$ monomers with positions $x_i(t)$ at time $t$. In the bead-spring model, with $k$ the bond's stiffness, we have
\begin{equation}
	\gamma \, \dot{x}_{i}=-k \sum_{j=1}^N M_{ij} x_{j} + f_i (t), \hspace{1cm} 1\le i\le N,
\label{LangevinAtt}
\end{equation}
$\gamma$ the friction coefficient on each monomer, and $f_i(t)$ stands for a centered Gaussian white noise, satisfying $\left \langle f_i(t) f_j(t') \right \rangle = 2 \gamma k_B T \delta(t-t') \delta_{ij}$. When the monomer of index $0$ is attached to the origin ($x_0=0$) and that the the other end is free, the connectivity matrix $M$ reads
\begin{equation}
M=
\begin{pmatrix}
2 & -1 & 0 & ... & ... & ... \\
-1 & 2 & -1 & 0 & ...  & ... \\
0 & -1 & 2 & -1 & ... & ... \\\
... & ...&...&...&...&... \\
... & ... &0 & -1 & 2 &-1 \\
... & ... & ... & 0 & -1 & 1 \\
\end{pmatrix}.
\end{equation}
We define $\tau_0=\zeta/k$ and $l_0=\sqrt{k_B T/k}$ which are respectively the typical relaxation time and the typical  length of one bond. In the following we use time and length units so that $\tau_0=1=l_0$. Note that in the literature the typical unit length is the three-dimensional bond length $b=\sqrt{3}l_0$ \cite{cao2015large}.

Here we aim to calculate the average first time $\langle \tau\rangle$ at which the last monomer $x_N$ reaches the threshold value $z$. We assume that $z\gg  \sqrt{N}$ so that reaching $z$ is a rare event. Following a scaling argument by De Gennes, we note that the motion of the end-monomer at $t$ involves a number $n(t)\propto \sqrt{t}$ of monomers (this can be seen by considering the non-noisy terms of Eq.~(\ref{LangevinAtt}) as a diffusion equation). Hence, after a fully extended configuration, in the rare event limit, the number of monomers involved in the dynamics of $x_N(t)$ is small compared to $N$ and we can consider the chain to be infinite (but discrete). Now we set a new index $m=N-n$ starting at zero for the free extremity, with $y_m=z-x_{N-m}$, and we consider now the average dynamics following an initial configuration which is at equilibrium conditional to $x_N=z$, for which $\langle x_n(0)\rangle=z n/N$ (i.e. $y_m(0)=m z/N$). In Laplace space, this dynamics reads
\begin{align}
\langle\tilde{y}_{m+1}(s)\rangle_z-(s+2)\langle \tilde{y}_m(s)\rangle_z+\langle\tilde{y}_{m-1}(s)\rangle_z=-m z/N
\end{align}
The characteristic polynomial of this recurrence equation is $P(r)=r^2-(s+2)r+1$ and has one negative and one positive root. Leaving aside the negative root term (unphysical because leading to infinite bond length for large $m$), and using the free end condition $y_1(t)=y_0(t)$, we obtain for the end monomer
\begin{align}
\langle \tilde{x}_N(s)\rangle_z=\langle \tilde{y}_0(s)\rangle_z= \frac{z}{2N}  \frac{ \sqrt{s+4}-\sqrt{s}  }{ s^{3/2}} 
\end{align} 
Taking the inverse Laplace transform leads to
\begin{align}
	&m_s(t)=\langle z-x_N(t)\rangle_z=\frac{\vert F\vert}{2k_BT}\psi(t), \hspace{2cm} \vert F\vert/k_BT=z/N \\
	&\psi(t)=  -1+e^{-2t}[(1+4t)I_0(2t)+4t I_1(2t) ], \label{psiRouse}
\end{align}
where $I_n$ represents the modified Bessel function of the first kind, and the previously found relation between $m_s(t)$ and the MSD $\psi(t)$ was used. We can check that
\begin{align}
\psi(t)\sim\begin{cases}
2t & (t\to0),\\
4\sqrt{t/\pi} & (t\to\infty),
\end{cases}
\end{align}
in agreement with a diffusive behavior at short times and a subdiffusive behavior at larger ones, where the value of the transport coefficient agrees with that of a monomer attached to a semi-infinite chain, see e.g. Ref.~\cite{Guerin2013}.

\begin{figure}%
\includegraphics[width=8cm]{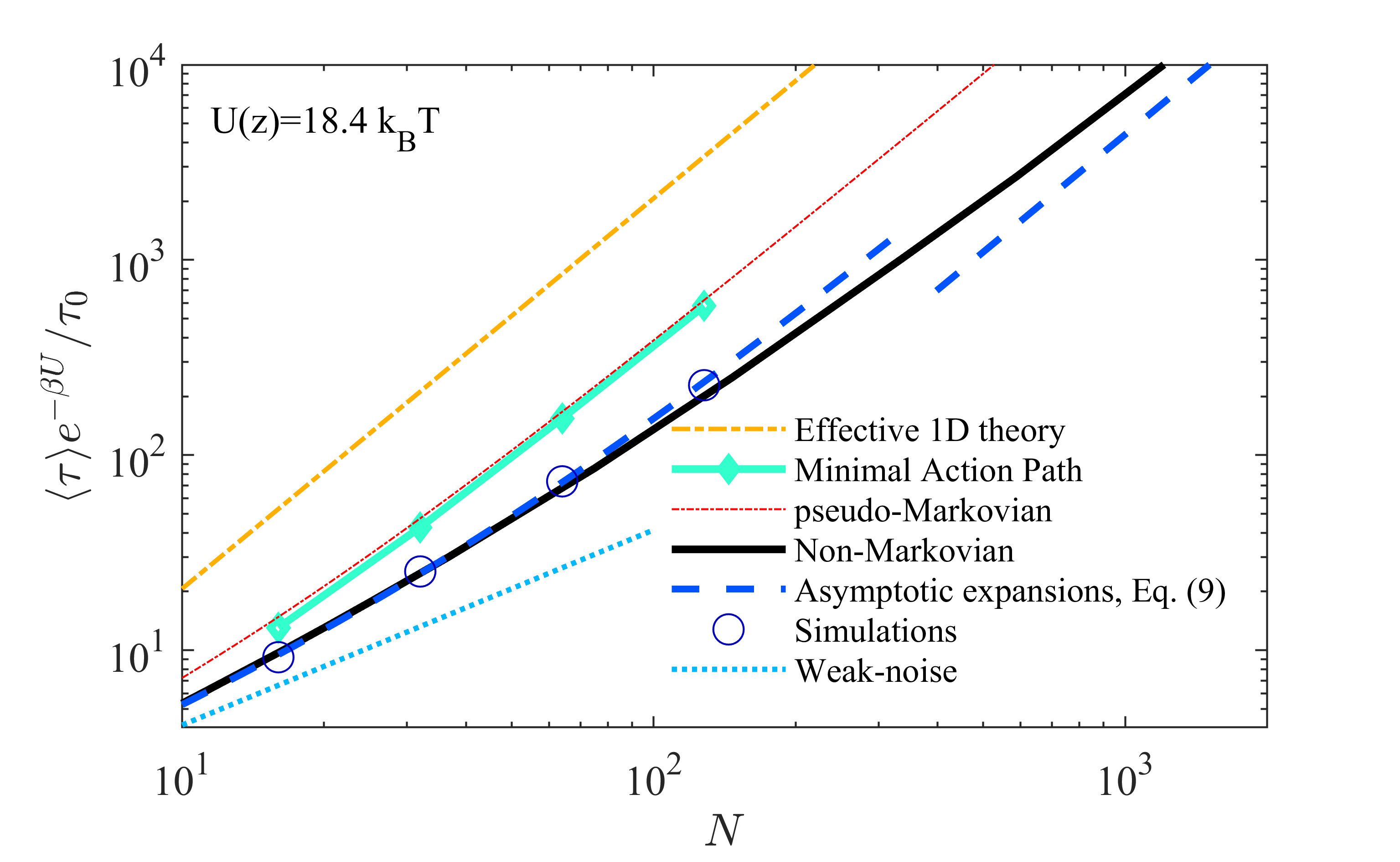}
\caption{Summary of theoretical approaches for the mean FPT for a flexible chain to reach an extension $z$, for $N=128$ monomers. Symbols are the simulations of Ref.~\cite{cao2015large}. The MAP theory corresponds to the Minimal Action Map method of Ref.~\cite{cao2015large}. The effective one dimensional approach is that of Milner and McLeish [Eq.~(\ref{Effective1D})]. The non-Markovian theory is evaluated by using Eq.~(\ref{EqTAppendix}). The Wilemski-Fixman approximation results from Eq.~(\ref{WilemskiFixman}). The weak-noise limit (at fixed $N$) corresponds to Eq.~(\ref{WeakNoise}). 
}
\label{figCompCao}%
\end{figure}

\subsection{Review of existing theories }
Here we review existing theoretical approaches to determine We explain how we obtain the curves on Fig.2 in the main text, which we reproduce here for clarity (Fig.\ref{figCompCao}). Note that a lot of these approaches are detailed in Ref.~\cite{cao2015large} and are mentioned here for completeness.

Historically, the present FPT problem was considered by Milner and McLeish for their theory of reptation in star polymer melts  \cite{milner1997parameter,milner1998reptation}. There they  approximated the dynamics of the whole chain by an effective Markovian dynamics for an attached dimer, with an effective friction coefficient $\gamma_e$. Once this approximation is made, the FPT problem can be solved exactly \cite{Redner:2001a,VanKampen1992}. The result in the regime $z\gg \sqrt{N}$ is  given by Eq.~(\ref{1DRes}) 
\begin{align}
\langle \tau\rangle \frac{e^{- z^2/2N}}{[2\pi  N]^{1/2}} =  \frac{\gamma_e N}{k z}.  \label{KRE}
\end{align}
In this approach, the effective friction has to be chosen. The choice of Milner and McLeish \cite{milner1997parameter,milner1998reptation} is   $\gamma_e=N\gamma/2$, so that in our units we have
\begin{align}
\langle \tau\rangle \frac{e^{- z^2/2N}}{[2\pi  N]^{1/2}} =  \frac{  N^2}{2  z}, \hspace{2cm}(\text{Effective one-dimensional approach)}\label{Effective1D}
\end{align}

Another important approach in reaction kinetics involving polymers is the Wilemski Fixman pseudo-Markovian approximation, where one assumes that all internal degrees of freedom are at equilibrium at the first passage. In the rare event limit, this can be implemented within our framework by setting $ \mu(t)=m_s(t)$, so that
\begin{align}
\langle \tau\rangle \frac{e^{- z^2/2N}}{[2\pi  N]^{1/2}} = \int_0^\infty dt\ \frac{e^{-F^2\psi(t)/8(k_BT)^2}}{[2\pi\psi(t)]^{1/2}}, \hspace{1cm}(\text{pseudo-Markovian/Wilemski-Fixman approximation)} \label{WilemskiFixman}
\end{align}
Another approach consists in calculating the fluctuations around minimal action paths~\cite{cao2015large}. Interestingly, this approach gives almost the same results as the Wilemski-Fixman approximation, meaning that both theories share similar hypotheses. 

Finally, in the weak noise limit (i.e., $T\to0$ at fixed  $\gamma,k,z$ and $N$), one can derive an explicit expression for the mean FPT as a function of the local first and second derivatives of the multidimensional potential $U(\{x_i\})=\sum k (x_{i+1}-x_i)^2/2$ (with a similar status to the Langer's theory for the passage through a multidimensional saddle point). For the Rouse chain, this expression was simplified by Cao et al.\cite{cao2015large} who showed that one obtains Eq.~(\ref{KRE}) with $\gamma_e=\gamma$, 
\begin{align}
\langle \tau\rangle \frac{e^{- z^2/2N}}{[2\pi  N]^{1/2}} =  \frac{  N}{  z}, \hspace{2cm}(\text{Weak-noise limit at fixed $N$)}\label{WeakNoise}
\end{align}
This result is exact in the limit of small noise, which here translates to $z\gg Nl_0$, while rare events kinetics can be assumed as soon as $z\gg \sqrt{N}l_0$. Note also that the same expression can be obtained by applying the general formula for the mean time out of non-characteristic domain [Eq.~(10.117) in the book \cite{schuss2009theory}] in the weak-noise limit.  
 
\section{Kinetics of wormlike chain closure} 
\label{SectionWLC_SI}

\subsection{Dynamics at the vicinity of a closed configuration}

\subsubsection{Average motion after a closure event}
Assuming overdamped motion in a viscous solvent without hydrodynamic interactions, the dynamics of a chain configuration $\ve[r](s,t)$  reads~\cite{hallatschek2007tension}
\begin{equation}
\boldsymbol \zeta\cdot\partial_t \ve[r] =-\kappa_b \ve[r]'''' + (\sigma \   \ve[r]' )'+\ve[f](s,t)\label{EqMotion1},
\end{equation}
where the prime denotes the derivative with respect to the curvilinear coordinate along the chain $s$, and  the tension $\sigma(s,t)$ is a Lagrange multiplier associated to the inextensibility constraint $(\ve[r]')^2 =1$. In the resistive force theory, the components of the friction tensor $\boldsymbol \zeta$ are $\zeta_{\parallel} $ and $\zeta_{\perp}$ in the directions that are respectively parallel and perpendicular to the local tangent vector, with $\zeta_{\perp}\simeq2\zeta_{\parallel}$ \cite{powers2010dynamics,DoiEdwardsBook}. Next, thermal forces satisfy
$\langle f_i(s,t)f_j(s',t')\rangle=2k_BT \zeta_{ij} \delta(s-s')\delta(t-t')$. We furthermore assume that no force and no torque conditions at the chain ends, where  ~\cite{liverpool2005dynamics,powers2010dynamics,hallatschek2007tension}
\begin{align}
\sigma \ve[r]'-\kappa_b \ve[r]'''=\ve[r]''=0
\end{align}
Note that combining these conditions with the inextensibility condition imply that the tension $\sigma$ vanishes at the extremities. 
 
We characterize now the average motion following a closure event for the WLC. Such motion is localized near the closing configurations of minimal bending energy. Such configurations are planar, let us call their local orientation $\phi$ in this plane. They are obtained by minimizing the functional
\begin{align}
\mathcal{F}=\int_0^L ds \frac{\kappa_b}{2}(\partial_s\phi)^2+F\left(\int_0^Lds\cos\phi-a\right),
\end{align}
where $F$ is the Lagrange multiplier associated to the constraint $\ve[r]_\mathrm{ee}=a\ve[e]_x$. Other Lagrange multipliers for the directions $y$ and $z$ could be included but would vanish at the end of the calculation. Minimizing $\mathcal{F}$ leads to
\begin{align}
\kappa_b\partial_s^2\phi^*(s)+F\sin\phi^*=0.
\end{align}
The full function $\phi^*(s)$ is needed to compute the equilibrium probability $p_s(r)$, which has been characterized elsewhere. Here we focus on the first passage kinetics and we will only need the behavior of $\phi^*$ near the ends. Using $\partial_s\phi^*\vert_{s=0}=0$, we obtain
\begin{align}
\phi^*(s)\simeq \phi_e-\frac{F\sin\phi_e\  s^2}{2\kappa_b} \hspace{2cm}(s\to0) \label{BehaviorPhiStar}
\end{align}
where $\phi_e=\phi^*(0)=\pi/2-\alpha$ is the initial angle in the closed configuration.  

Now, we consider the dynamics near a chain extremity  when initial conditions are equilibrium closed configurations, first in the absence of noise. We quantify the lateral motion (with respect to the local orientation at the extremity at closure) by using the ansatz 
\begin{align}
\ve[r](s,t)\simeq \ve[u]^*(0)s+\ve[n]^*(0)r^\perp(s,t),
\end{align}
where $\ve[u]^*(0)$ is the orientation of the optimal configuration at the extremity and $\ve[n]^*(0)$ a unit vector perpendicular to $\ve[u]^*(0)$. Inserting this ansatz into Eq.~(\ref{EqMotion1}) (with vanishing thermal forces) and projecting in the direction $\ve[u]^*(0)$ we obtain $\sigma'=0$. Since the tension vanishes at the ends, we conclude that it is negligible near the ends, for $s>0$. The dynamics in the lateral direction then reads
\begin{align}
\zeta_\perp\partial_t r^\perp=-\kappa_b \partial_s^4 r^\perp. \label{DynLin}
\end{align}
Denoting by $\phi_e$ the orientation of $\ve[u]^*(0)$, we measure the small deviations of the orientation at later times by 
\begin{align}
\phi=\phi_e+\phi_1+..., \hspace{2cm} \phi_1=\partial_s r_\perp.
\end{align}
The local orientation is thus solution of 
\begin{align}
\zeta_\perp\partial_t \phi_1=-\kappa_b \partial_s^4 \phi_1, \label{EqPhi1}
\end{align}
and the initial and boundary conditions lead to 
\begin{align}
&\phi_1(s,t=0)=-\frac{F\sin\phi_e s^2}{2\kappa_b} \label{IniPhi1},\\
&\partial_s\phi_1(s=0,t)=\partial_s^2\phi_1(s=0,t)=0 \label{BCPhi1}
\end{align}
%Importantly, the initial condition (\ref{IniPhi1}) do not satisfy the boundary (\ref{BCPhi1}). This apparent paradox is in fact the sign  that the deformation is local, and can be solved by looking solutions of the form
%\begin{align}
%\phi_1(s,t)=\varepsilon(t) A(s/\varepsilon(t))
%\end{align}
%with $A$ a scaling function
 
We take the Laplace transform of Eq.~(\ref{EqPhi1})  with respect to the time $t$, with $p$ the Laplace variable:
  \begin{align}
\zeta_\perp\left(p\tilde{\phi_1}+\frac{F\sin\phi_e s^2}{2\kappa_b}\right)=-\kappa_b \partial_s^4\tilde{\phi}_1(s,p). 
\end{align}
The solution of this equation which satisfies all boundary conditions (and does not diverge exponentially at $s\to\infty$) is
\begin{align}
\tilde{\phi}_1(p,s)=-\frac{F\sin\phi_e }{\kappa_b}
\left\{\frac{s^2}{2p}
+\left(\frac{\kappa_b}{\zeta_\perp}\right)^{1/2}\frac{e^{-s\left(\frac{\zeta_\perp p}{4\kappa_b}\right)^{1/4}}}{p^{3/2}}\left(\cos\left[s\left(\frac{\zeta_\perp p}{4\kappa_b}\right)^{1/4} \right]+\sin\left[s\left(\frac{\zeta_\perp p}{4\kappa_b}\right)^{1/4} \right]
\right) \right\}\label{PHI1}
\end{align}
The unit tangent vector $\ve[u](s,t)$ reads
\begin{align}
\ve[u](s,t)=\cos\phi \ve[e]_1+\sin\phi\ve[e]_2\simeq \ve[u]^*(0)-(\sin \phi_e)\phi_1\ve[e]_1+(\cos\phi_e)\phi_1 \ve[e]_2+\mathcal{O}(\phi_1^2)
\end{align}
Using $\ve[u]=\partial_s\ve[r]$, we see that 
\begin{align}
\partial_s x_1(s,t)=-\sin\phi_e \phi_1(s,t). \label{Diffx}
\end{align}
where $x_1=\ve[r]\cdot\ve[e]_1$. Using the argument that for fixed $t$, the chain remains underformed for large $s$, we get  
\begin{align}
x_1(s,t)-x_1(s,0)=+\sin\phi_e \int_s^\infty [\phi_1(u,t)-\phi_1(u,0)]. 
\end{align}
Consider now $X_1(t)=x(s=L,t)-x(s=0,t)$, we get 
\begin{align}
\langle \Delta X_1(t) \rangle \equiv \langle X_1(t)-X_1(0)\rangle= -2\sin\phi_e \int_0^\infty du \ [\phi_1(u,t)-\phi_1(u,0)].
\end{align}
Going to Laplace space and using (\ref{PHI1}), we obtain
\begin{align}
\langle \tilde{\Delta X}_1(p)\rangle=2\sqrt{2}  \frac{(\sin\phi_e)^2 F}{(\kappa_b p^7 \zeta_\perp^3)^{1/4}}. 
\end{align}
Finally, inverting the Laplace transform leads to 
\begin{align}
\langle X_1(t)-X_1(0)\rangle=\frac{2\sqrt{2}(\sin\phi_e)^2 F }{\Gamma(7/4)(\kappa _b \zeta_\perp^3)^{1/4}}t^{3/4}.
\end{align}
%For a monomer at the interior, the above relation holds if one divides the numerical coefficient by a factor of 4. 

\subsubsection{Identification of the subdiffusion coefficient}

Now, we add the fluctuations, so that Eq.~(\ref{DynLin}) becomes
\begin{align}
\zeta_\perp\partial_t r^\perp=-\kappa_b \partial_s^4 r^\perp+f(s,t), \hspace{2cm}\langle f(s,t)f(s',t')\rangle=2k_BT\zeta_\perp\delta(t-t')\delta(s-s'),
\end{align}
which is associated to the boundary conditions (\ref{BCPhi1}). From the work presented in Ref.~\cite{guerin2014semiflexible}, we can extract
\begin{align}
\text{Var}(r_1^\perp(s=0,t))=\frac{2\sqrt{2}k_BT}{\Gamma(7/4)\kappa_b^{1/4}}\left(\frac{t}{\zeta_\perp}\right)^{3/4} \label{FIE2}.
\end{align}
where the notation $\text{Var}(...)$ represents the variance. 
With $X_1(t)=\sin\phi_e[r_1^\perp(L,t)-r_1^\perp(0,t)]$ and $X_2(t)=\cos\phi_e[r_1^\perp(L,t)+r_1^\perp(0,t)]$, we see that $X_1,X_2$ and $X_3$ are independent, and that
\begin{align}
 \mathrm{Var}(X_1(t))= \kappa(\sin\phi_e)^2 t^{3/4},\hspace{0.2cm}
\mathrm{Var}(X_2(t))=  \kappa(\cos\phi_e)^2 t^{3/4},\hspace{0.2cm}
\mathrm{Var}(X_3(t))=  \kappa \ t^{3/4},\hspace{0.2cm}
 \kappa=\frac{4\sqrt{2}k_BT}{\Gamma(7/4)\kappa^{1/4}}\left(\frac{1}{\zeta_\perp}\right)^{3/4}.
\end{align}
It is important to note that  the relation
 \begin{align}
\frac{\mathrm{Var}(X_1(t))}{\langle X_1(t)-X_1(0)\rangle}=\frac{2k_BT}{F} 
\end{align}
holds, which is consistent with the interpretation of $F=-\partial_r E^*(r)$ as an effective force applied from $t>0$. 

As a supplementary check, we may compare our Eq.~(\ref{FIE2}) to other results from the literature. The subdiffusion coefficient is given for an interior monomer  in Ref.~\cite{bullerjahn2011monomer}
\begin{align}
\text{Var}(r^\perp(s=0,t))=\frac{k_BT}{2\sqrt{2}\Gamma(7/4)\kappa^{1/4}}\left(\frac{t}{\zeta_\perp}\right)^{3/4}.
\end{align}
In Ref.~\cite{guerin2014semiflexible} it was shown that for exterior monomers the coefficient has to be multiplied by 4 [see Eq.~(29) in Ref.~\cite{guerin2014semiflexible}], but doing so leads to a subdiffusion coefficient that is twice smaller than in Eq.~(\ref{FIE2}). However, we think that this discrepancy comes from a typo in Ref.~\cite{bullerjahn2011monomer}.  Indeed, there the authors show that the displacement of an interior monomer at small times satisfies
\begin{align}
r^\perp(t)-r^\perp(0)=\frac{1}{2\sqrt{2}\Gamma(3/4)\kappa_b^{1/4}\zeta_{\perp}^{3/4}}\int_0^t\frac{\theta(\tau)}{(t-\tau)^{1/4}}, \hspace{2cm}\langle \theta(t)\theta(t')\rangle=\frac{2\sqrt{2}k_BT\kappa_b^{1/4}\zeta_\perp^{3/4}}{\Gamma(1/4)\vert t-t'\vert^{3/4}},
\end{align}
see their Eqs.~(6) and (10). Hence, the Mean Square Displacement reads
\begin{align}
\langle[r^\perp(t)-r^\perp(0)]^2\rangle= \frac{k_BT}{2\sqrt{2}\Gamma(3/4)^2\Gamma(1/4)\kappa_b^{1/4}\zeta_{\perp}^{3/4}}  \times 2 \int_0^tds\int_0^sds'\frac{1}{\vert s-s'\vert^{3/4}(t-s)^{1/4}(t-s')^{1/4}},
\end{align}
where a factor of two comes from the fact that we evaluated the integral for $s'<s$ only. Performing the integral leads to
\begin{align}
\langle[r^\perp(t)-r^\perp(0)]^2\rangle= \frac{k_BT \ t^{3/4}}{ \sqrt{2}\Gamma(7/4) \kappa_b^{1/4}\zeta_{\perp}^{3/4}} 
\end{align}
which is two times larger than the result indicated after Eq.~(10) in Ref~\cite{bullerjahn2011monomer}. Multiplying by 4  \cite{guerin2014semiflexible} we recover Eq.~(\ref{FIE2}), confirming the validity of our analysis.  

\subsection{FPT analysis in the regime $a \gg L^2/l_p$}
\begin{figure}%
\includegraphics[width=8cm]{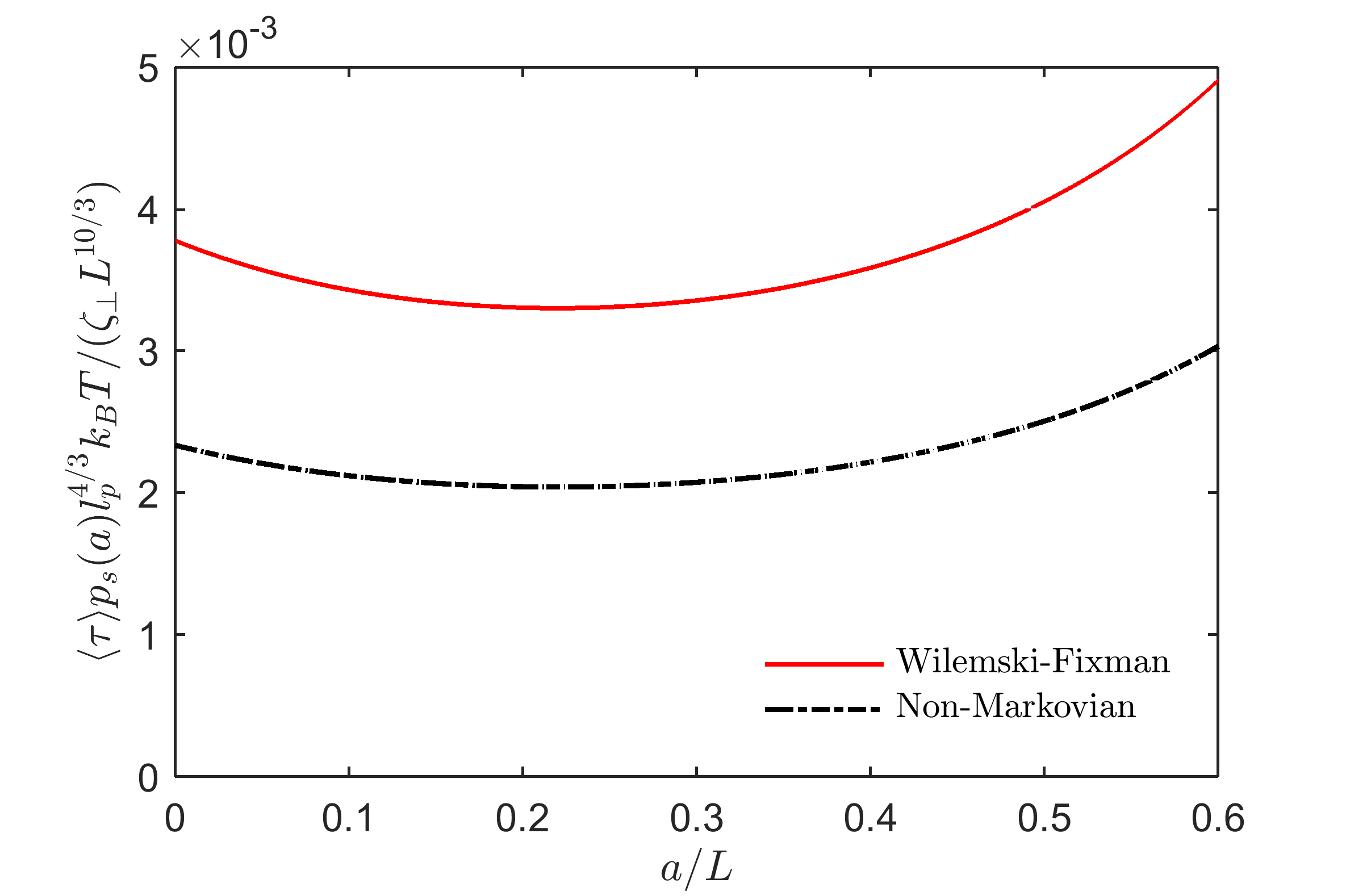}
\caption{Predictions of the non-Markovian theory for the mean FPT of a wormlike chain to reach an end-to-end distance smaller than $a$, in the limit $a l_p/L^2\to\infty$ at fixed $a/L$. 
}
\label{figFPT_WLC_largea}%
\end{figure}

In the  limit $a\gg L^2/l_p$ (while still keeping the small capture radius condition $a\ll L$), we remark that the MSD becomes comparable to $\langle X_1(t) \rangle^2$ at length scales of the order of  $L^2/l_p$. Since we assume $a\gg L^2/l_p$, we can thus assume the $X_i(t)$ are small compared to $a$ at these relevance length scales, so that the end-to-end distance is $r_\mathrm{ee}=[(a+X_1)^2+X_2^2+X_3^2]^{1/2}\simeq a + X_1$. This means that we can consider the motion as being one-dimensional along the direction $\ve[e]_1$, and we can apply the formalism presented above with $H=3/8$. We directly apply  
(\ref{T_AH}), with a subdiffusion coefficient $\kappa(\sin\phi_e)^2$, obtaining 
\begin{align}
\langle \tau\rangle \ p_s(a) =  \frac{A_{3/8}}{ (\kappa\sin^2\phi_e)^{4/3}(\beta F)^{5/3}}
\end{align}
We may write
\begin{align}
\langle \tau\rangle \ p_s(a) = \frac{\zeta_\perp L^{10/3}}{k_BT l_p^{4/3}} \left(\frac{\Gamma(7/4)}{4\sqrt{2} } \right)^{4/3} \frac{A_{3/8}}{ ( \sin^2\phi_e)^{4/3}\lambda^{5/3}}
\end{align}
Our best estimate of the constant $A_H$ for $H=3/8$ is $A_{3/8}=2.1$ (which is obtained by numerically solving (\ref{EqMuH}). This is about $1.6$ times smaller than its estimate in the Wilemski-Fixman approximation $A_{3/8}^{\mathrm{WF}}=3.39$ . Here, we have defined $\lambda=-\partial_{\tilde{r}} E^*(\tilde{r})$, where $E^*$ is the dimensionless minimal bending energy to form a loop of size $\tilde{r}=a/L$ (the true minimal bending energy is $\mathcal{E}_b=\kappa_b E^*/L$. Hence the above formula predicts the value of the kinetic prefactor as a function of the angle $\phi_e$ and the local derivative of the minimal bending energy, such quantities are well known from the analysis of equilibrium configurations. We represent on Fig.~\ref{figFPT_WLC_largea} the value of this kinetic prefactor as a function of $a/L$. We see that it does not vary much as long as $a<0.6 L$ (above this value we cannot really speak of closure anyway). Hence, we may approximate it by its value for $a\ll L$, while still $a\gg L^2/l_p$, for which we obtain with $\lambda=21.55$ and $\varphi_e\simeq2.281$ (radians): 
\begin{align}
\langle \tau\rangle \ p_s(a) = 0.0023 \frac{\zeta_\perp L^{10/3}}{k_BT l_p^{4/3}}  
\end{align}
In the Wilemski-Fixman approach, we obtain with $A_{3/8}^{\mathrm{WF}}=3.39$
\begin{align}
\langle \tau\rangle \ p_s(a) = 0.0037 \frac{\zeta_\perp L^{10/3}}{k_BT l_p^{4/3}}  \label{LimitingWF}
\end{align}

\subsection{The Wilemski-Fixman approximation}
\begin{figure}%
\includegraphics[width=8cm]{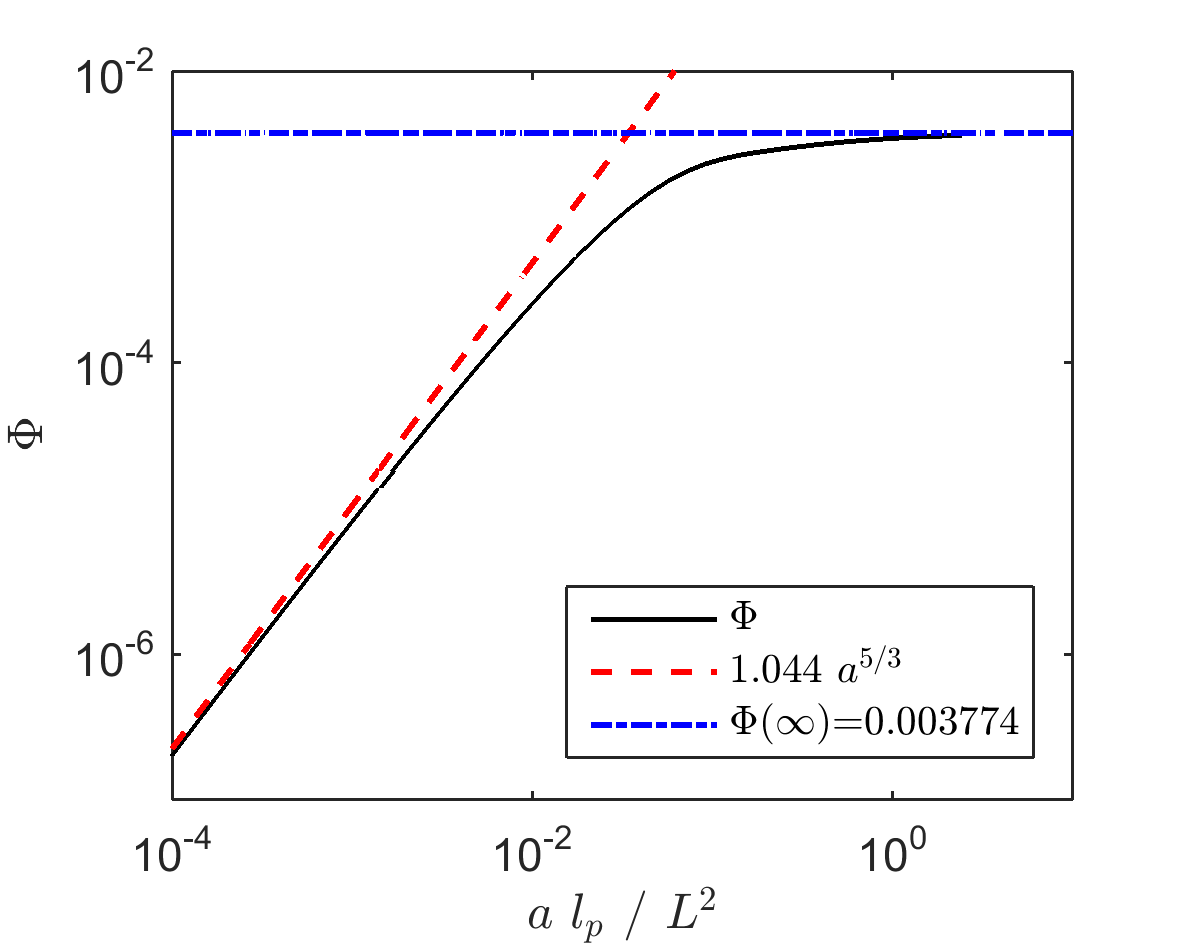}
\caption{ Scaling function $\Phi$ appearing in Eq.~(\ref{ScalingTWF}) for the mean FPT of wormlike chains in the Wilemski-Fixman approximation for $a\ll L$. Limiting behaviors [Eqs. (\ref{LimitingWF}),(\ref{Small_aWF})] are also indicated.  
}
\label{figScalingPHI}%
\end{figure}

In the Wilemski-Fixman approximation, we can write
\begin{align}
\langle \tau\rangle \ p_s(a)=\int_0^\infty dt \ p(a,t\vert a,0)\label{IntegralWF}
\end{align}
where $p_s(r)$ is the pdf of the end-to-end distance $r$, and $p(a,t\vert a,0)$ is the probability density of $r=a$ at $t$ given that at $t=0$ we have equilibrium conditions conditional to $r=a$. This probability can be identified by projecting the three-dimensional Gaussian dynamics of $X_i(t)$: 
\begin{align}
p(a,t\vert a,0)=\int d\ve[X]\int d\ve[X]_0 \delta\left(a-(X_1^2+X_2^2+X_3^2)\right) p(\ve[X],t\vert \ve[X]^{(0)}0,0) p_s(\ve[X]_0\vert r=a) 
\end{align} 
Here, $p(\ve[X],t\vert \ve[X]^{(0)},0)$ is the Gaussian propagator for $\ve[X]=(X_1,X_2,X_3)$, which reads
\begin{align}
p(\ve[X],t\vert \ve[X]^{(0)},0)=&\frac{1}{\vert\cos\phi_e\sin\phi_e\vert (2\pi\kappa t^{3/4})^{3/2}}\times \nonumber\\
&\exp\left\{-\frac{1}{2\kappa t^{3/4}}\left[\frac{\left(X_1-X_1^{(0)}-\frac{F\sin^2\phi_e \kappa t^{3/4}}{2k_BT}\right)^2}{ \sin^2\phi_e }+\frac{\left(X_2-X_2^{(0)}\right)^2}{ \cos^2\phi_e }+ \left(X_3-X_3^{(0)}\right)^2  \right]\right\}\label{propag}
\end{align}
and $p_s(\ve[X]_0\vert r=a)$ is the equilibrium distribution of $\ve[X]$ given that $r=a$:
\begin{align}
p_s(\ve[X]^{(0)} \vert r=a)= \frac{1}{\mathcal{Z}} e^{ \beta F X_1^{(0)}} \delta\left(\vert\ve[X]^{(0)}\vert-a\right)\label{EqCondPs}.
\end{align}
with the normalization
\begin{align} 
\mathcal{Z}=\int d\ve[X]^{(0)} e^{\beta F X_1^{(0)}} \delta\left(\vert\ve[X]^{(0)}\vert-a\right)=\frac{4 \pi a \sinh(a \beta F)}{\beta F}\label{EqZ}.
\end{align}
In these expressions, we remind that
\begin{align}
\kappa=\frac{4\sqrt{2}}{\Gamma(7/4)} l_p^{1/4}\left(\frac{k_BT}{\zeta_\perp}\right)^{3/4}, \hspace{1cm} \beta F\simeq\lambda_0\frac{l_p}{L^2}.
\end{align}
with the numerical values \cite{shimada1984} $\lambda_0\simeq21.55$ and $\varphi_e\simeq2.281$ radians. All terms appearing in the the integral (\ref{IntegralWF}) giving the mean FPT in the Wilemski-Fixman approximation are thus identified. 
We now realize that the change of variables $t\to \tilde{t}=t/t^*$, with $t^*=\zeta_\perp L^{16/3}/[l_p^{7/3}k_BT]$ leads to the scaling behavior 
\begin{align}
\langle \tau\rangle \ p_s(a)= \frac{\zeta_\perp L^{10/3}}{k_BT l_p^{4/3}}\Phi\left(\tilde{a}\right),\ \hspace{1cm}\tilde a=a l_p/L^2 \label{ScalingTWF}.
\end{align} 
The function $\Phi$ can be evaluated numerically by evaluating the integral   (\ref{IntegralWF}) with the use of (\ref{propag}), (\ref{EqCondPs}) (\ref{EqZ}), in which we evaluate all terms by setting $1=l_p=L=\zeta_\perp=k_BT$ and by using spherical coordinates. The resulting function $\Phi$ is represented on Fig.~\ref{figScalingPHI}.

We now derive the limiting behaviors of $\Phi$. For small $\tilde{a}$, the limiting value of $\Phi$ can be obtained by performing the change of variable $X_i\to X_i/a$, $X_i^{(0)}\to X_i^{(0)}/a$ in which case we realize that in this limit the baising force $F$ is irrelevant and can be taken as zero. We find 
\begin{align}
&\Phi(\tilde a\to 0) \simeq  c \ \tilde{a}^{5/3},\nonumber\\
& c  =\int_0^\infty dt \int d\ve[X]\int d\ve[X]^{(0)} 
\frac{\delta(\vert \ve[X]\vert-1)\delta(\vert \ve[X]^{(0)}\vert-1)}{4\pi \vert\cos\phi_e\sin\phi_e\vert (2\pi\kappa_0 t^{3/4})^{3/2}}  
\ e^{ -\frac{1}{2\kappa_0 t^{3/4}}\left[\frac{\left(X_1-X_1^{(0)} \right)^2}{ \sin^2\phi_e }+\frac{\left(X_2-X_2^{(0)}\right)^2}{ \cos^2\phi_e }+ \left(X_3-X_3^{(0)}\right)^2  \right] }   \simeq 1.044 \label{Small_aWF}
\end{align} 
The opposite limit $a\gg L^2/l_p$, the closure time in the Wilemski-Fixman approximation is given  by  Eq.~(\ref{LimitingWF}).

\end{document}